%% file: paper.tex
\documentclass[submission, Phys]{SciPost}

\input{incl_settings.tex} 
\input{incl_shortcuts.tex}

\graphicspath{{./figs/}}

\begin{document}

\begin{center}{\Large \textbf{\color{scipostdeepblue}{
MadNIS at NLO
}}}
\end{center}

\begin{center}
Giovanni De Crescenzo\textsuperscript{1},
Javier Mariño Villadamigo\textsuperscript{1},\\
Nina Elmer\textsuperscript{2},
Theo Heimel\textsuperscript{3},
Tilman Plehn\textsuperscript{1,4},
Ramon Winterhalder\textsuperscript{5},
Marco Zaro\textsuperscript{5}
\end{center}

\begin{center}
{\bf 1} Institut f\"ur Theoretische Physik, Universit\"at Heidelberg, Germany
\\
{\bf 2} DAMTP, University of Cambridge, Cambridge, United Kingdom\\
{\bf 3} CP3, Universit\'e catholique de Louvain, Louvain-la-Neuve, Belgium
\\
{\bf 4} Interdisciplinary Center for Scientific Computing (IWR), Universit\"at Heidelberg, Germany \\
{\bf 5} TIFLab, Universit\`a degli Studi di Milano \& INFN Sezione di Milano, Italy
\end{center}

\begin{center}
\today
\end{center}

\section*{\color{scipostdeepblue}{Abstract}}
{\bf 
We combine fast amplitude surrogates with neural importance sampling to accelerate NLO calculations. For virtual corrections, a learned ratio to the Born matrix element with calibrated uncertainties achieves high precision across phase space. For real emission, we stick to the standard FKS subtraction and train sector-conditioned surrogates of the regularized integrands away from divergences. MadNIS then uses multi-channel mappings and FKS sectors as conditions. We validate our approach for electron-positron scattering to three and four jets and find significant speed-ups and variance reduction in the integration.
}

\vspace{3pt}
\noindent\rule{\linewidth}{1pt}
\tableofcontents\thispagestyle{fancy}
\noindent\rule{\linewidth}{1pt}
\vspace{3pt}

\clearpage
%%%%%%%%%%%%%%%%%%%%%%%%%%%%%%%%%%%%%%%%%%%%%%%%%%%
\section{Introduction}
\label{sec:introduction}

Precise and scalable event generation is the key theme in theoretical particle physics~\cite{Campbell:2022qmc}, as the upcoming High-Luminosity LHC (HL-LHC) will push complexity and luminosity to unprecedented levels. Event generators such as \pythia~\cite{Bierlich:2022pfr}, \sherpa~\cite{Sherpa:2024mfk}, \herwig~\cite{Bellm:2025pcw}, and \texttt{MadGraph}~\cite{Maltoni:2002qb,Alwall:2007st,Alwall:2011uj}, specifically \mg~\cite{Alwall:2014hca,Frederix:2018nkq}, provide the backbone of the first-principles simulation chain, combining perturbative QCD calculations with parton showers and hadronization. Together with the subsequent detector simulation, they allow us to compare precise predictions with measured data. In data science these events would be referred to as digital twins, and the comparison with measured data as simulation-based inference.

Next-to-leading order (NLO) and even higher order predictions are essential for precision LHC physics, but their complex phase-space integrations and repeated evaluations of expensive matrix elements constitute a major computational bottleneck. The precision and simulation statistics required by the HL-LHC implies a rapidly growing computational cost and motivates the use of modern machine learning (ML)~\cite{Plehn:2022ftl,Ubiali:2026myh} to accelerate all components of the simulation pipeline~\cite{Butter:2022rso} and the simulation workflow~\cite{Plehn:2026gxv}. 

Neural networks have been shown to speed up amplitude calculations~\cite{Bishara:2019iwh,Badger:2020uow,Aylett-Bullock:2021hmo,Maitre:2021uaa,Danziger:2021eeg,Winterhalder:2021ngy,Janssen:2023ahv,Maitre:2023dqz,Brehmer:2024yqw,Breso-Pla:2024pda,Herrmann:2025nnz,Favaro:2025pgz,Villadamigo:2025our,Bahl:2026jvt} including a correctly calibrated uncertainty estimate~\cite{Badger:2022hwf,Bahl:2024gyt,Bahl:2025xvx,Bahl:2026qaf, Beccatini:2025tpk}, improve hadronization~\cite{Ilten:2022jfm,Ghosh:2022zdz,Chan:2023ume,Bierlich:2023zzd,Chan:2023icm,Bierlich:2024xzg,Assi:2025avy,Butter:2025wxn}, generate complete collider events~\cite{Hashemi:2019fkn,DiSipio:2019imz,Butter:2019cae,Alanazi:2020klf,Butter:2023fov,Butter:2021csz,Quetant:2024ftg}, and accelerate detector simulations~\cite{Paganini:2017hrr,Erdmann:2018jxd,Buhmann:2020pmy,Krause:2021ilc,Krause:2021wez,Buhmann:2021caf,Chen:2021gdz,Diefenbacher:2023vsw,Xu:2023xdc,Diefenbacher:2023flw,Ernst:2023qvn,Hashemi:2023rgo,Favaro:2024rle,Buss:2024orz,Krause:2024avx}. Supplementing these various surrogates, neural importance sampling~\cite{Bendavid:2017zhk,Klimek:2018mza,Chen:2020nfb,Gao:2020vdv,Deutschmann:2024lml} has been successfully applied at leading order (LO) using \madnis~\cite{Heimel:2022wyj,Heimel:2023ngj,Heimel:2024wph} or its \sherpa counterpart~\cite{Gao:2020zvv, Bothmann:2020ywa, Bothmann:2025lwg}. Normalizing-flow samplers have also been used at NLO~\cite{Gao:2020zvv} and NNLO accuracy in multi-jet final states~\cite{Janssen:2025zke}. 

A unified NLO implementation of ultrafast amplitude surrogates and neural importance sampling is the natural next step in ML-enhanced event generation. Theory predictions beyond LO require evaluating Born, virtual, and integrated subtraction amplitudes for the Born-like phase space, together with real and subtraction terms for the real-emission phase space. The soft, collinear, and soft--collinear singularities are regularized by a suitable subtraction scheme~\cite{Catani:1996vz,Catani:2002hc,Frixione:1995ms,Frederix:2009yq}. This structure provides a substantial challenge for a combined ML-surrogate and sampling strategy.

In this first study, we show how to combine learned amplitude surrogates with neural importance sampling to accelerate fixed-order NLO calculations, while preserving the standard subtraction structure.
We employ the FKS scheme, where the real-emission contribution is decomposed into sectors labeled by an FKS parton-sister pair. Building on this structure, we provide an NLO version of the \madnis framework. 
For virtual corrections, we find that learning the ratio of the subtracted virtual correction to the Born matrix element provides the best balance between speed and precision.
The learned uncertainty is calibrated across phase space and quantifies the surrogate precision.
For real emission, we develop surrogates for the finite FKS-sector cross sections, treating the FKS sector as a discrete label, and use them only where no local subtraction is active.
Using a conditioning on these FKS labels in addition to the standard conditioning on the multi-channels allows us to combine the virtual and real surrogates with the \madnis sampling of the Born-like and real-emission phase space. 
As a controlled benchmark, we consider $e^+e^- \to 3$ and 4 jets at NLO QCD, where all infrared singularities are in the final state and only a few partonic channels contribute.
This is a first step towards LHC processes, which require initial-state subtraction terms and many more partonic channels, as discussed in Sec.~\ref{sec:outlook}.

The paper is structured as follows: Section~\ref{sec:fks_basic} reviews the FKS subtraction, Sec.~\ref{sec:surrogates} introduces the amplitude surrogates, Sec.~\ref{sec:sampling} combines them with \madnis, and Sec.~\ref{sec:perf} shows the precision and acceleration. We conclude with and outlook in Sec.~\ref{sec:outlook}.

\clearpage
%%%%%%%%%%%%%%%%%%%%%%%%%%%%%%%%%%%%%%%%%%%%%%%%%%%
\section{FKS subtraction recap}
\label{sec:fks_basic}

To establish our notation, we consider the generic scattering process,
\begin{align}
  p_a + p_b \;\to\; p_1 + p_2 + \cdots + p_n\; .
\end{align}
Its NLO correction consists of $n$-particle (Born-like) and $(n\!+\!1)$-particle (real emission) final states. We write the NLO cross section as
\begin{align}
    \sigma_\text{NLO} = \int_n \left[\dd\sigma^\text{B} + \dd\sigma^\text{V}\right] + \int_{n+1} \dd\sigma^\text{R} \; .
\label{eq:nlo_def}
\end{align}
Over the Born-like phase space, we evaluate the Born contribution and the virtual corrections. The real-emission corrections are defined over the $(n\!+\!1)$-particle phase space. While $\sigma_\text{NLO}$ is infrared-finite~\cite{Kinoshita:1962ur,Lee:1964is}, the Born-like and real-emission integrals are individually divergent. Numerically, we regularize each integral using a subtraction term,
\begin{align}
    \sigma_\text{NLO} 
    &= \sigma_n + \sigma_{n+1} \notag \\
    &\equiv \int_n \left[\dd\sigma^\text{B} + \dd\sigma^\text{V} + \dd\sigma^\text{I} \right] + \int_{n+1} \left[\dd\sigma^\text{R}-\dd\sigma^\text{S}\right]
    \qquad \mwith \qquad \dd\sigma^\text{I}=\int_1 \dd\sigma^\text{S}\;.
\label{eq:subtraction_def}
\end{align}
The $(n\!+\!1)$-particle subtraction term $\dd\sigma^{\text{S}}$ is constructed such that it has the same local divergences as $\dd\sigma^\text{R}$ and its integral $\dd\sigma^\text{I}$ analytically cancels the corresponding divergence in $\dd\sigma^\text{V}$, leaving only finite terms. 
We denote these finite remainders by 
$\dd\sigma^V_\text{fin}$ and $\dd\sigma^I_\text{fin}$, with
\begin{align}
\dd\sigma^\text{V} + \dd\sigma^\text{I} = \dd\sigma^\text{V}_\text{fin} + \dd\sigma^\text{I}_\text{fin} \; .
\end{align}
That way, both integrals become finite and can be implemented in a numerical Monte Carlo generator.
Away from the singular limits, the subtraction term $\dd\sigma^\text{S}$ can be chosen freely, and different choices define different subtraction schemes.
We employ the FKS subtraction scheme~\cite{Frixione:1995ms,Frederix:2009yq}, which splits the real-emission phase space into FKS sectors for all possible pairs of particles that can introduce soft, collinear or soft--collinear singularities in the real matrix element.

%%%%%%%%%%%%%%%%%%%%%%%%%%%%%%%%%%%%%%%%%%%%%%%%%%%
\subsubsection*{Born-like contributions}

Following Eq.\eqref{eq:subtraction_def}, the first term of the Born-like cross section is the leading order contribution
\begin{align}
    \dd\sigma^\text{B} = \frac{1}{2s {\mathcal{N}_{n}}} \, \mata^\text{B}(\Phi_{n}) \,\dd\Phi_n
    \qquad
    \mwith \qquad 
    \Phi_{n}=(p_1,\dots,p_n)\;,
    \label{eq:born_contrib}
\end{align}
where $\mata^\text{B}$ denotes the averaged squared Born matrix element, $\mathcal{N}_{n}$ the symmetry factor for identical particles in the final state, $s$ the squared center-of-mass energy, and $\dd\Phi_n$ the phase-space element. 
The finite virtual contribution to the cross section arises from the interference of the one-loop and Born amplitudes,
\begin{align}
    \dd\sigma^\text{V}_\text{fin}=\frac{1}{2s \mathcal{N}_n} \, \mata^\text{V}(\Phi_{n}) \,\dd\Phi_n \; ,
    \label{eq:virtual_contrib}
\end{align}
where $\mata^\text{V}(\Phi_n)$ denotes the finite part of the one-loop interference term, evaluated in conventional dimensional regularization, which regularizes both ultraviolet and infrared divergences, as defined, for instance, in App.~B of Ref.~\cite{Frederix:2009yq}.
Finally, we write the finite contribution of the integrated subtraction term as
\begin{align}
    \dd\sigma^\text{I}_\text{fin}=\frac{\mata^\text{I}(\Phi_{n})}{2s \mathcal{N}_n} \, \dd\Phi_n
    \quad \mwith \quad
    \mata^\text{I}(\Phi_{n})=\frac{\alpha_s}{2\pi}\mathcal{Q}(\Phi_{n})\,\mata^\text{B}(\Phi_{n}) + \frac{\alpha_s}{2\pi}\sum_{k,l}\mathcal{E}_{kl}(\Phi_{n})\,\mata^\text{B}_{kl}(\Phi_{n})\;,
\end{align}
where $\mata^\text{B}_{kl}$ denotes the color-linked Born amplitudes, and $\mathcal{Q}$ and $\mathcal{E}$ are the finite parts of the integrated subtraction term~\cite{Frederix:2009yq}.
The combined Born-like contribution then reads
\begin{align}
\sigma_n 
= \int \dd\Phi_n \, \frac{1}{2s\mathcal{N}_{n}}  \left[ \mata^\text{B}(\Phi_{n})+\mata^\text{V}(\Phi_{n})+\mata^\text{I}(\Phi_{n}) \right] \equiv \int \dd\Phi_n \, f_n(\Phi_n) \;.
\label{eq:def_n}
\end{align}
%

%%%%%%%%%%%%%%%%%%%%%%%%%%%%%%%%%%%%%%%%%%%%%%%%%%%
\subsubsection*{Real emission}

The real-emission phase space extends the Born kinematics $\Phi_n$ by additional radiation variables that parametrize the soft and collinear limits,
\begin{align}
    \xi_i = 2\frac{E_i}{\sqrt{s}}
    \qqqquad
    y_{ij} = \cos\theta_{ij} = \frac{\mathbf p_i\!\cdot\!\mathbf p_j}{|\mathbf p_i|\,|\mathbf p_j|}
    \qqqquad
    \varphi_i = \text{azimuthal angle} \; .
    \label{eq:rad_variables}
\end{align}
For each radiated parton $i$ and FKS partner $j$, the FKS sector function $\mathcal S_{ij}(\Phi_n,\xi_i,y_{ij},\varphi_i)$ isolates the singular region associated with the pair $(i,j)$ while suppressing all others. The sector functions are normalized such that the phase-space volume is preserved, \ie
\begin{align}
    \sum_{ij}\mathcal S_{ij}(\Phi_n,\xi_i,y_{ij},\varphi_i)=1 \; .
\end{align}
In a given FKS sector $ij$, we then define the regularized sector amplitude
\begin{align}
    \Sigma_{ij}(\Phi_n,\xi_i,y_{ij},\varphi_i)
    =
    (1-y_{ij})\,\xi_i^2\,
    \mata^\text{R}(\Phi^{(ij)}_{n+1})\,
    \mathcal S_{ij}(\Phi_n,\xi_i,y_{ij},\varphi_i)\;.
    \label{eq:sigma}
\end{align}
The multiplicative prefactor regularizes the averaged squared real-emission matrix element $\mata^\text{R}$ in the soft and collinear limits of the selected sector, where $\Phi^{(ij)}_{n+1}$ is constructed from the underlying Born configuration $\Phi_n$ and the radiation variables $\Phi^{ij}_\text{rad}\equiv(\xi_i,y_{ij},\varphi_i)$. The quantity $\Sigma_{ij}$ is related to the one denoted by the same symbol in Ref.~\cite{Frederix:2009yq}, but is not identical to it, as we do not include the phase-space factor. The singular soft, collinear, and soft--collinear configurations are obtained by taking the corresponding limits of the radiation variables, namely $\xi_i\to 0$ for the soft limit and $y_{ij}\to 1$ for the collinear limit. This defines the relevant real-emission phase-space configurations
\begin{alignat}{3}
\Phi_{n+1}^{\text{hard}} &\equiv \Phi^{(ij)}_{n+1}
\qquad\qquad
& \Phi_{n+1}^{\text{soft}} &\equiv \Phi^{(ij)}_{n+1}\Big|_{\xi_i=0} \notag\\
\Phi_{n+1}^{\text{coll}} &\equiv \Phi^{(ij)}_{n+1}\Big|_{y_{ij}=1}
\qquad
& \Phi_{n+1}^{\text{soft--coll}} &\equiv \Phi^{(ij)}_{n+1}\Big|_{\xi_i=0,\,y_{ij}=1} \, .
\end{alignat}
In the soft and soft--collinear limits, these configurations coincide kinematically with the underlying Born configuration. The phase-space construction is discussed in more detail in Sec.~\ref{sec:sampling}.
The fully subtracted real-emission contribution can then be written as
\begin{align}
\sigma_{n+1}
=
\sum_{ij}
\int \dd \Phi^{(ij)}_{n+1}\,
\frac{1}{2s}
\frac{\mathcal{A}^{\text{R-S}}_{ij}(\Phi^{(ij)}_{n+1})}{\mathcal{N}_{n+1}}
\equiv
\sum_{ij}
\int \dd \Phi^{(ij)}_{n+1}\,
f^{\,ij}_{n+1}(\Phi^{(ij)}_{n+1}) \; ,
\label{eq:fks_expand}
\end{align}
with
\begin{align}
\mata_{ij}^{\text{R}-\text{S}}(\Phi^{(ij)}_{n+1})
= \frac{1}{\xi_i^2 (1-y_{ij})} \;\Bigg[
&\Sigma_{ij}(\Phi_n,\xi_i,y_{ij},\varphi_i) \notag \\
-&\left|\frac{\partial\Phi^{\text{coll}}_{n+1}}{\partial\Phi^{\text{hard}}_{n+1}}\right|\,\Sigma_{ij}(\Phi_n,\xi_i,1,\varphi_i)\,\Theta(y_{ij}-1+\delta)\ \notag \\
-&\left|\frac{\partial\Phi^{\text{soft}}_{n+1}}{\partial\Phi^{\text{hard}}_{n+1}}\right|\,\Sigma_{ij}(\Phi_n,0,y_{ij},\varphi_i)\,\Theta(\xi_\text{cut}-\xi_i) \notag \\
+&\left|\frac{\partial\Phi^{\text{soft--coll}}_{n+1}}{\partial\Phi^{\text{hard}}_{n+1}}\right|\,\Sigma_{ij}(\Phi_n,0,1,\varphi_i)\,\Theta(y_{ij}-1+\delta)\,\Theta(\xi_\text{cut}-\xi_i)\Bigg] \; .
\label{eq:real_sub}
\end{align}
The terms in brackets consist of the locally regularized real-emission contribution together with its collinear, soft, and soft--collinear subtraction terms, each weighted by the corresponding phase-space factor.
The parameters $\xi_\text{cut}$ and $\delta$ define the regions in which the subtractions are active. Physical predictions combining  Born-like and real-emission contributions are formally independent of these parameters, but they can have an impact on the efficiency of the numerical integration. 
We initially stick to the default choice in \mg, namely
\begin{align}
\xi_\text{cut} = 0.5
\qqquad \text{and} \qqquad 
\delta = 1 \; .
\label{eq:default_cutoffs}
\end{align}
With this choice, the subtraction terms are active over a comparatively large fraction of the real-emission phase space.
This reduces the fraction of negative weights and the variance of the integral, but it also enlarges the region with large cancellations between real emission and subtraction terms, where surrogates are hard to use.

%\clearpage
%%%%%%%%%%%%%%%%%%%%%%%%%%%%%%%%%%%%%%%%%%%%%%%%%%%
\section{Amplitude surrogates}
\label{sec:surrogates}

Learned amplitude surrogates are our first key ingredient for ultrafast NLO calculations. 
We refer to the squared matrix elements and interference terms $\mata$ defined in Sec.~\ref{sec:fks_basic} simply as amplitudes.
As a benchmark process, we consider jet production in $\Pep \Pem$ annihilation. While surrogate models for tree-level matrix elements will only lead to major efficiency gains for large jet multiplicities, a substantial acceleration of the virtual contributions appears within reach.
Here, the virtual corrections account for $\sim95\%$ ($3$-jet) and $\sim98\%$ ($4$-jet) of the
integrand evaluation time.

\begin{figure}[t]
    \includegraphics[width=0.30\textwidth]{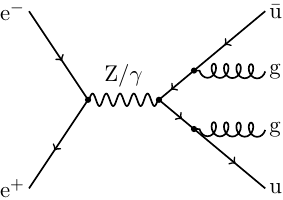}
    \hspace{5mm}
    \includegraphics[width=0.30\textwidth]{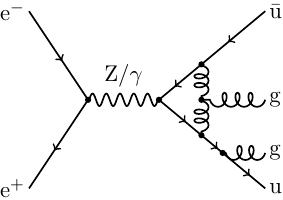}
    \hspace{5mm}
    \includegraphics[width=0.30\textwidth]{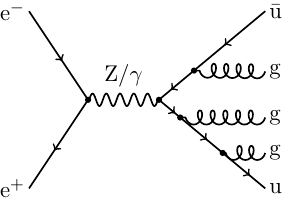}
    \caption{Left to right: representative Feynman diagrams for the Born, virtual, and real-emission contributions to the NLO prediction for the $\Pep \Pem \to \Pu \Pubar \Pg \Pg$ process.}
    \label{fig:diagrams_4_jet}
\end{figure}

We assume a center-of-mass energy of $\sqrt{s}=1\,\tev$ and restrict ourselves to a subset of representative partonic subprocesses at leading order, 
\begin{align}
  \text{3-jet (Born)}\qqquad
  &\Pep \Pem \to \Pu \Pubar \Pg \notag\\
  \text{4-jet (Born)}\qqquad
  &\Pep \Pem \to \Pu \Pubar \Pg \Pg \; .
\label{eq:def_procs_lo}
\end{align}
Since the infrared structure is identical for all massless quark flavors, we focus on up quarks. The NLO QCD corrections include virtual corrections and the real-emission subprocesses
\begin{align}
  \text{3-jet (real)}\quad
  &\Pep \Pem \to \Pu \Pubar \;\Pg \Pg  \notag\\
  &\Pep \Pem \to \Pu \Pubar \;\Pq \Pqbar  \notag\\[1ex]
  \text{4-jet (real)}\quad
  &\Pep \Pem \to \Pu \Pubar \;\Pg \Pg \Pg  \notag\\
  &\Pep \Pem \to \Pu \Pubar \;\Pg \Pq \Pqbar 
  \qquad \text{where} \qquad \Pq=\Pu,\Pd,\Pc,\Ps \; .
\label{eq:def_procs_nlo}
\end{align}
Illustrative Feynman diagrams for the Born, virtual, and real-emission contributions to the 4-jet process are shown in Fig.~\ref{fig:diagrams_4_jet}.
Using the 3-jet case for illustration, with analogous considerations applying to the 4-jet case, we highlight some aspects of the singularity structure: 
\begin{itemize}
\item For $\Pep \Pem \to \Pu \Pubar \;\Pg \Pg $,
there exist five collinear configurations, since each gluon can become collinear to the quark or antiquark, or the two gluons can form a collinear pair. We employ the symmetry over gluon exchange to reduce the number of sectors down to 3, which we denote as sectors 1, 2, and 3.

\item In the case of $\Pep \Pem \to \Pu \Pubar \;\Pq \Pqbar $ and assuming $\Pq \neq \Pu$, two collinear singularities appear: $\Pq \| \Pqbar$, with the corresponding Born term $\Pep \Pem \to \Pu \Pubar \Pg$, and $\Pu \| \Pubar$ with the Born process $\Pep \Pem \to \Pq \Pqbar \Pg$. We focus on the former, as the latter is suppressed by the FKS function $\mathcal{S}$, which results in two sectors we denote by 4 (gluon splitting to a down-like quark pair) and 6 (gluon splitting to a $\Pc \Pcbar$ pair).
\item For the same real emission and $\Pq = \Pu$, each $\Pu$ can become collinear with either $\Pubar$, which gives 4 singular configurations.
Like in the first bullet, these singular configurations are symmetric (under quark or antiquark exchange) and only one of them is independent, giving rise to sector 5.
\end{itemize}
Altogether, we have to take into account six FKS sectors. They can be written in terms of the underlying potentially divergent emission,
\begin{align}
  &\text{Sector 1:}\quad \Pu \to \Pu \Pg 
  \qqqquad &\text{Sector 2:}\quad \Pubar \to \Pubar \Pg
  \qqqquad &\text{Sector 3:}\quad \Pg \to \Pg \Pg \notag\\
  &\text{Sector 4:}\quad \Pg \to \Pd \Pdbar (\Ps \Psbar)
  \qqqquad &\text{Sector 5:}\quad \Pg \to \Pu \Pubar
  \qqqquad &\text{Sector 6:}\quad \Pg \to \Pc \Pcbar\; . 
\label{eq:3-jet-sectors}
\end{align}
%

%------------------------------------------
%------------------------------------------

%%%%%%%%%%%%%%%%%%%%%%%%%%%%%%%%%%%%%%%%%%%%%%%%%%%
\subsection{Born-like surrogates}
\label{sec:nbody_reg}

As indicated in Eq.\eqref{eq:def_n}, we divide the Born-like contributions into $\mata^\text{B}$, $\mata^\text{V}$, and $\mata^\text{I}$. A network surrogate can encode individual contributions or the combined Born-like amplitude. We generate a set of external momenta with \madnis and train a regression network to learn the phase-space functions
\begin{alignat}{7}
    &\text{Partial sum} &
    \mata^\text{BV} &= \mata^\text{B} + \mata^\text{V} \notag \\
    &\text{Ratio V/B} &
    R^\text{V/B} &= \frac{\mata^\text{V}}{\mata^\text{B}} \notag \\
    &\text{Total sum} &
    \mata^\text{BVI} &= \mata^\text{B}+\mata^\text{V}+\mata^\text{I} \notag \\
    &\text{Ratio (VI)/B} & \qqquad 
    R^\text{(VI)/B} &= \frac{\mata^\text{V}+\mata^\text{I}}{\mata^\text{B}}\; .
    \label{eq:amp_sectors}
\end{alignat}
Because the integrated subtraction term $\mata^\text{I}$ contains logarithms of the subtraction parameters, in particular terms proportional to $\log\delta$ and $\log\xi_\text{cut}$, the corresponding regression targets inherit this dependence.
The quantities $\mata^\text{BVI}$ and $R^{\text{(VI)}/\text{B}}$ are therefore defined for the default values given in Eq.\eqref{eq:default_cutoffs}.
We use a fully connected multilayer perceptron (MLP) as the network architecture.
Providing the network with several parametrizations of the same phase-space point is known to improve training convergence and accuracy~\cite{Bahl:2024gyt}.
We therefore combine the final-state 4-momenta with the invariants, where we use logarithms because the invariants span several orders of magnitude
\begin{align}
    y^{\text{B}}=(\Phi_n, \log s^{\text{B}}_{kl}) \qquad \mwith \qquad 
    s^\text{B}_{kl} = p_k\cdot p_l \quad \mfor \quad k\ne l \;.
\end{align}
For the same reason, we either learn the logarithm of the amplitudes or an amplitude ratio
\begin{align}
    f_{\theta}(y^{\text{B}}) \approx f(y^{\text{B}})
    \qquad \mwith \qquad f \in \{\log \mata, R\} \; .
\end{align}
This gives us two ways to obtain the Born-like amplitudes.
We either learn them directly, or we multiply the learned ratio with the exact Born amplitude, which is cheap to evaluate,
\begin{alignat}{7}
  \mata_\theta^\text{BV} 
  \qqquad \text{vs} \qqquad 
    \mata_\theta^\text{BV} &\equiv R^\text{V/B}_\theta \times \mata^\text{B} + \mata^\text{B} \notag \\
  \mata_\theta^\text{BVI} 
  \qqquad \text{vs} \qqquad 
    \mata_\theta^\text{BVI} &\equiv R^\text{(VI)/B}_\theta \times \mata^\text{B} + \mata^\text{B} \; .
\label{eq:def_amps_2}
\end{alignat}  
The index $\theta$ marks the quantity encoded by the network.
We never learn $\mata^\text{V}$ alone, because it spans a wide range and changes sign.
In the ratio, regions with negative $\mata^\text{V}$ simply appear as negative values of $R^\text{V/B}$, as shown in Fig.~\ref{fig:target2}.

%------------------------------------------
\begin{figure}[t]
    \includegraphics[width=0.495\linewidth, page=1]{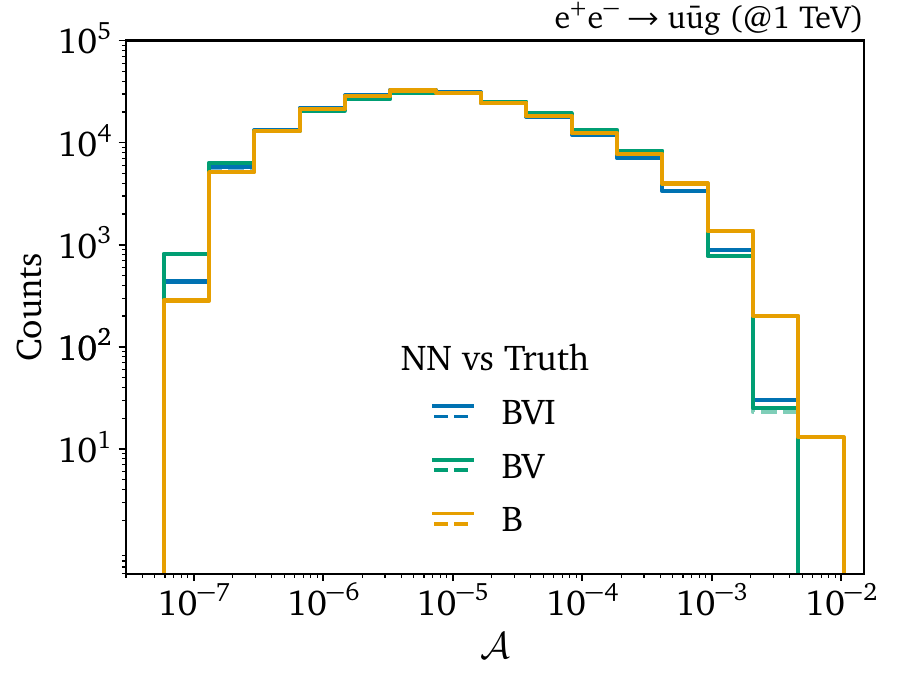} 
    \includegraphics[width=0.495\linewidth, page=2]{figs/nbody/preds_adapt.pdf}
    \caption{Learned Born, combination of Born and virtual contributions without the integrated subtraction term, and full Born-like amplitudes. We show results for 3-jet (left) and 4-jet (right) production. The solid lines indicate surrogates, the dashed lines the truth.
    }
    \label{fig:target1}
\end{figure} 
%------------------------------------------

Beyond accurate amplitude predictions, we also want a corresponding uncertainty.
We let the network learn its own uncertainty by giving it two outputs, the amplitude prediction $f_\theta$ and its variance $\sigma^2_{f,\theta}$, and training it with the heteroscedastic loss
\begin{align}
    \loss = \sum_{i=1}^{N_\text{data}}
    \left[ \frac{\left[f_i-f_\theta(y^{\text{B}}_i)\right]^2}{2\sigma^2_{f,\theta}(y^{\text{B}}_i)}+\log\sigma_{f,\theta}(y^{\text{B}}_i) \right] \; .
    \label{eq:het_loss}
\end{align}
It follows from maximizing a Gaussian likelihood with learned mean and variance~\cite{Plehn:2022ftl}.
This loss yields a stable mean and a calibrated uncertainty~\cite{Bahl:2024gyt,Bahl:2025xvx,Bahl:2026qaf}, and a detailed discussion appears in Ref.~\cite{Bahl:2025xvx}.
When learning the ratio, the Born amplitude is exact, so the learned amplitude uncertainty follows directly from the learned ratio uncertainty
\begin{align}
    \sigma_{\mata,\theta} = \mata^\text{B} \, \sigma_{R,\theta} \; .
\end{align}
When learning $\log \mata$ directly, the surrogate amplitude is $\mata_\theta = \exp f_\theta$, and linear error propagation gives
\begin{align}
\sigma_{\mata,\theta} = \mata_\theta \, \sigma_{f,\theta} \; ,
\end{align}
which is valid for $\sigma_{f,\theta} \ll 1$.
The network hyperparameters are listed in Tab.~\ref{tab:hyperparameter_MLP}.
In Fig.~\ref{fig:target1}, we show the learned and true Born, BV, and BVI amplitudes for the 3-jet (left) and 4-jet (right) processes.
They span roughly five and seven orders of magnitude, respectively.
At the level of these distributions, all surrogates agree well with the truth.
Figure~\ref{fig:target2} shows that both ratios are strongly peaked and limited in range, which makes them easy regression targets.

%------------------------------------------
\begin{figure}[t]
    \includegraphics[width=0.495\linewidth, page=3]{figs/nbody/preds_adapt.pdf}
    \includegraphics[width=0.495\linewidth, page=4]{figs/nbody/preds_adapt.pdf}
    \caption{Ratios of the learned amplitudes to the Born contribution, shown for the combined virtual and integrated subtraction term (left) and for the virtual contribution alone (right). The solid lines indicate the surrogates, the dashed lines the truth.}
    \label{fig:target2}
\end{figure} 
%------------------------------------------

%-------------------------------
\begin{figure}[b!]
    \includegraphics[width=0.495\linewidth, page=2]{figs/nbody/benchmark_uug_full.pdf}
    \includegraphics[width=0.495\linewidth, page=2]{figs/nbody/benchmark_uugg_full.pdf} \\
    \includegraphics[width=0.495\linewidth, page=1]{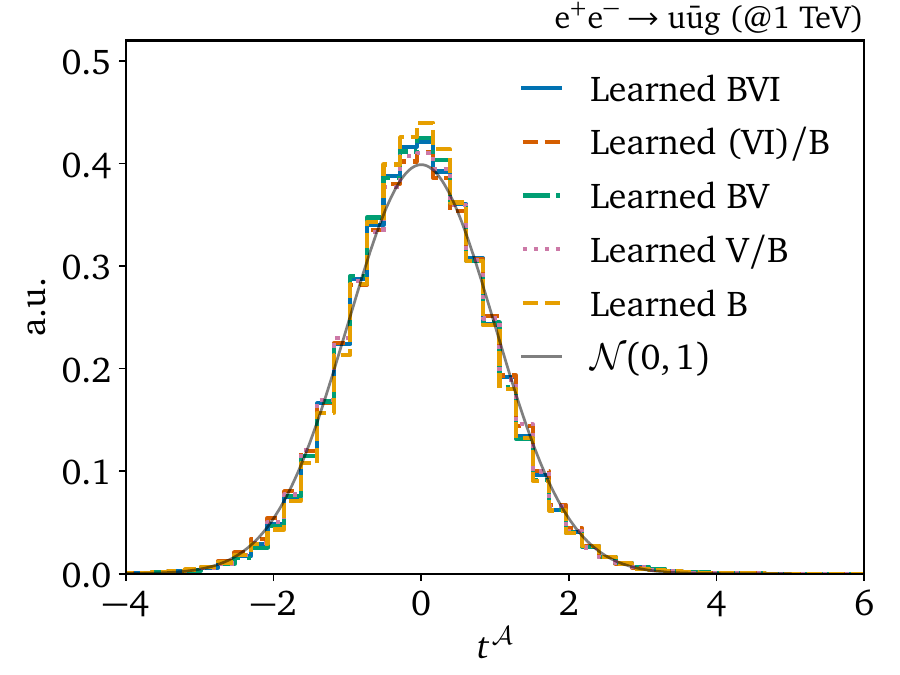} 
    \includegraphics[width=0.495\linewidth, page=1]{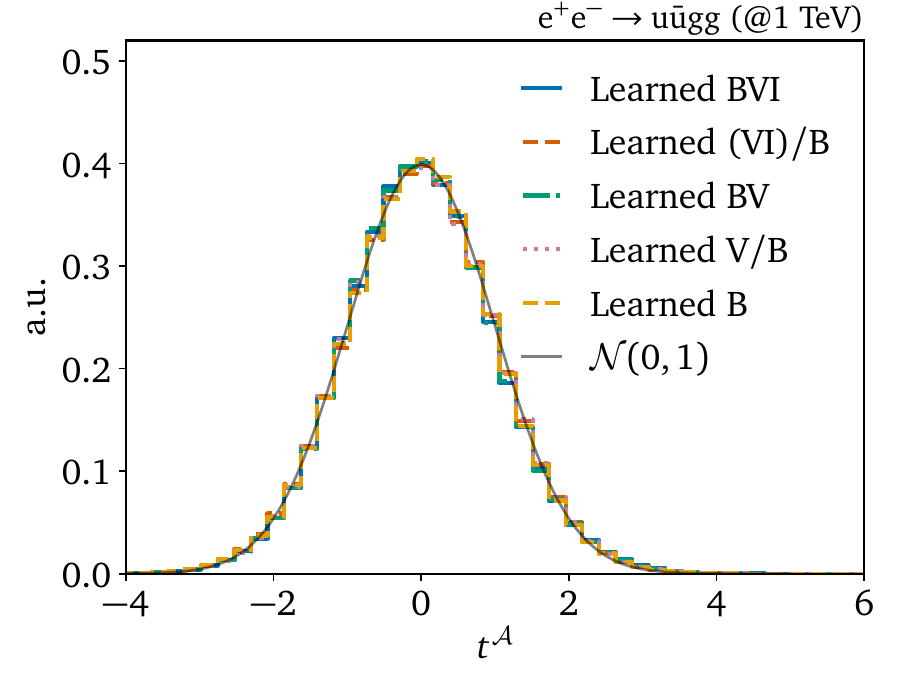}
    \caption{Relative accuracies for the virtual amplitudes defined in Eq.\eqref{eq:def_delta} (upper) and pull distributions defined in Eq.\eqref{eq:pull} (lower) for the different options. We show results for 3-jet (left) and 4-jet (right) production.}
    \label{fig:benchmark_nbody}
\end{figure} 
%------------------------------

To compare the learned amplitudes and the learned ratios, we evaluate the relative deviation from the truth over phase space
\begin{align}
    \Delta(y^{\text{B}}) =  \frac{\mata_\theta(y^{\text{B}}) - \mata(y^{\text{B}})}{\mata(y^{\text{B}})} \; .
    \label{eq:def_delta}
\end{align}
In the upper panels of Fig.~\ref{fig:benchmark_nbody}, we show the relative accuracies for the learned BV and BVI amplitudes using the different strategies.
Learning the amplitudes directly is clearly less accurate.
Learning the ratio and rescaling with the Born amplitude reduces the median relative deviation to around $10^{-6}$ for the 3-jet process and $10^{-4}$ for the 4-jet process.
For the 3-jet process, essentially no phase-space points deviate by more than one per mille, and for the 4-jet process hardly any deviate by more than one percent.
Both ratios perform well, and for the analysis in Sec.~\ref{sec:perf} we use the virtual-to-Born ratio $R_\theta^\text{V/B}$, which is slightly more accurate and  does not depend on the subtraction parameters we re-optimize in Sec.~\ref{sec:optimizing_subtraction}.

Finally, we test the calibration of the learned uncertainties with the pull over the same phase space
\begin{align}
    t(y^{\text{B}}) = \frac{\mata_\theta(y^{\text{B}})-\mata(y^{\text{B}})}{\sigma_{\mata,\theta}(y^{\text{B}})} \; .
    \label{eq:pull}
\end{align}
The heteroscedastic loss assumes Gaussian errors, so for an unbiased surrogate with calibrated uncertainty the pull follows a unit Gaussian $\mathcal{N}(0,1)$~\cite{Bahl:2024gyt,Bahl:2025xvx}.
The lower panels of Fig.~\ref{fig:benchmark_nbody} confirm this for all surrogates, which also validates the Gaussian error model.

%%%%%%%%%%%%%%%%%%%%%%%%%%%%%%%%%%%%%%%%%%%%%%%%%%%
\subsection{Real-emission surrogates}

%-------------------------------
\begin{figure}[b!]
\includegraphics[width=0.495\linewidth, page=1]{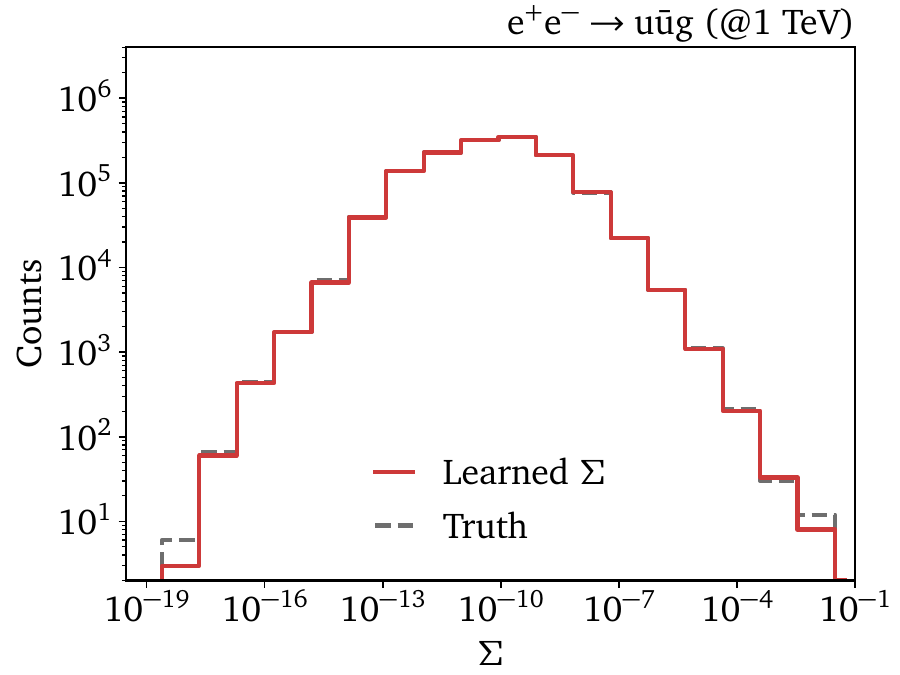}
\includegraphics[width=0.495\linewidth, page=1]{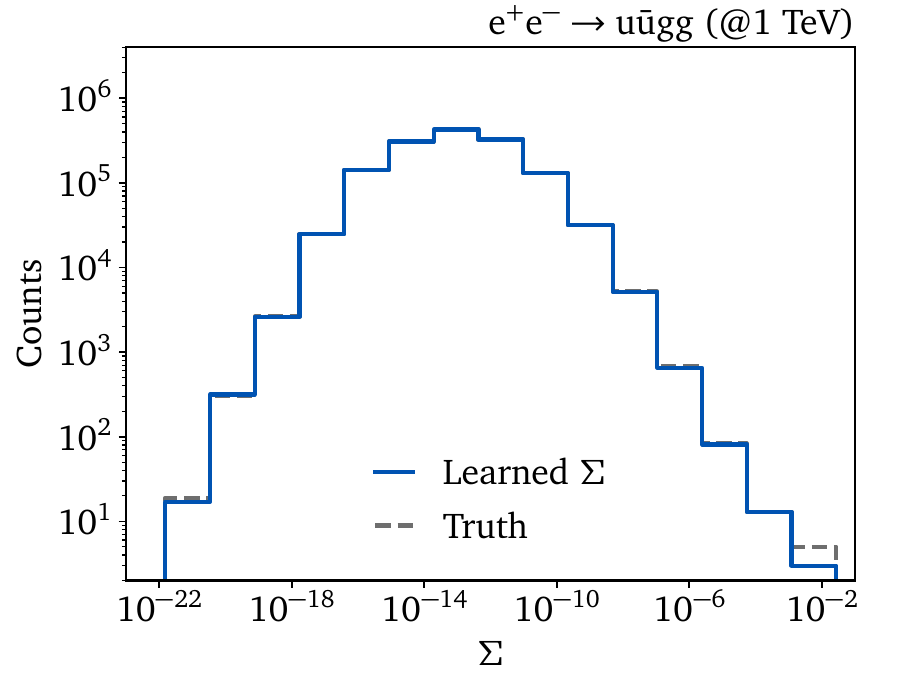}
\caption{Learned and true real-emission amplitudes for the NLO corrections to 3-jet (left) and 4-jet (right) production. The solid lines indicate the surrogates, the dashed lines the truth. We denote the target function $\Sigma$ without subscripts since we combine all FKS sectors.}
\label{fig:target_and_benchmark_sigma}
\end{figure} 
%-------------------------------

For real emission, the regression target is the regularized amplitude $\Sigma_{ij}$ defined in Eq.\eqref{eq:sigma}.
Preprocessing becomes even more important, because $\Sigma_{ij}$ spans a much wider range than the virtual correction.
The reason is that the FKS function $\mathcal{S}_{ij}$ suppresses all singular regions that do not belong to the pair $ij$
\begin{align}
    \mathcal{S}_{ij}(\Phi_n,\xi_i,y_{ij},\varphi_i) \to 0 \qquad \text{when} \qquad \mathbf{p}_k\parallel\mathbf{p}_l 
    \quad \text{or} \quad E_k=0 \qquad \text{for} \quad \{k,l\}\neq\{i,j\} \; .
\end{align}
This leads to arbitrarily small amplitudes, and the relevant range of $\Sigma$ covers more than 15 orders of magnitude, as illustrated by the 4.2M training amplitudes in Fig.~\ref{fig:target_and_benchmark_sigma}.

In addition, $\Sigma_{ij}$ is in general not Lorentz invariant, because $\mathcal{S}_{ij}$ depends on the energies and angles of the outgoing particles.
We therefore use the input
\begin{align}
 \left\{ \text{lin-}\log{s_{kl}^{\text{R}}},E_k,y_{kl},\varphi_k \right\} 
 \qqquad \mwith \qqquad 
 s_{kl}^{\text{R}}=p_k\cdot p_l 
 \qquad 
 y_{kl} = \cos{\theta_{kl}} \qquad (k\ne l) \;.
\end{align}
The invariants vanish exactly in singular configurations, where the logarithm is not defined.
We therefore apply a lin-log transformation, which is linear below a small threshold and logarithmic above it. We choose this threshold based on the training data such that only genuine singular kinematics fall in the linear regime. The threshold value and all other network hyperparameters are listed in Tab.~\ref{tab:hyperparameter_real_MLP}.

%-------------------------------
\begin{figure}[t]
\includegraphics[width=0.495\linewidth, page=1]{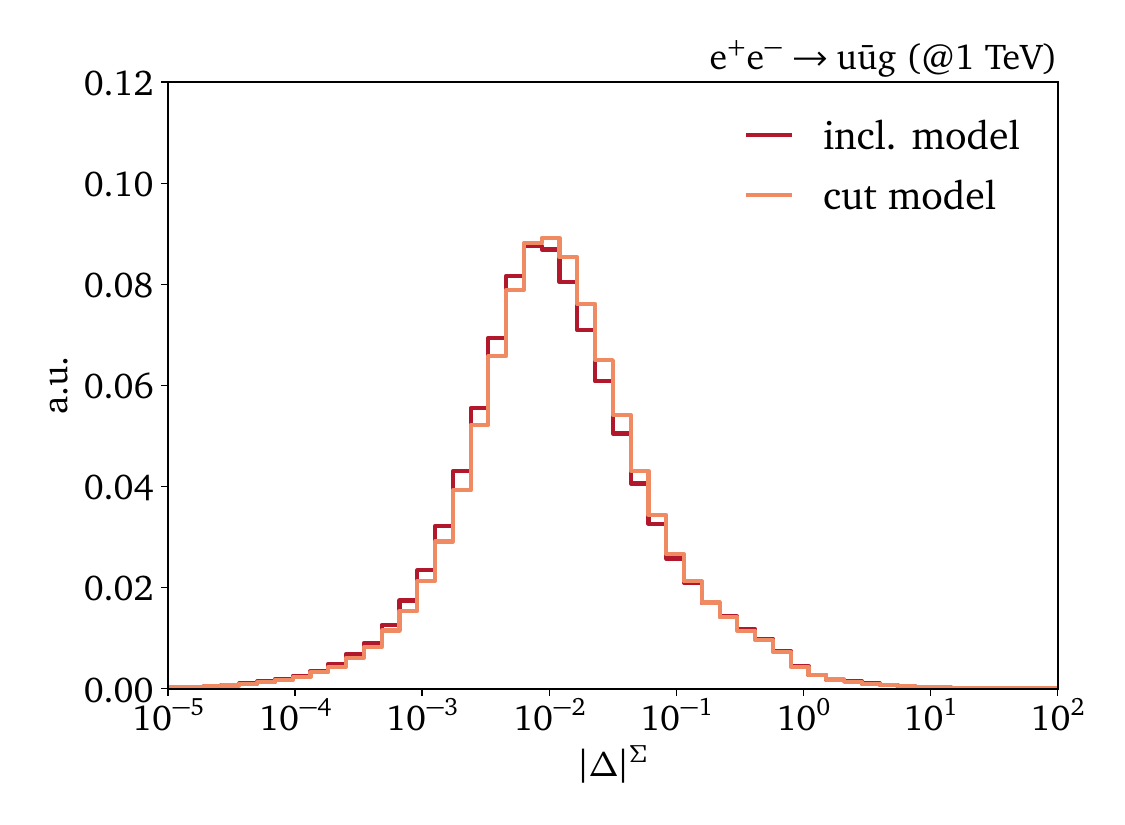}
\includegraphics[width=0.495\linewidth, page=1]{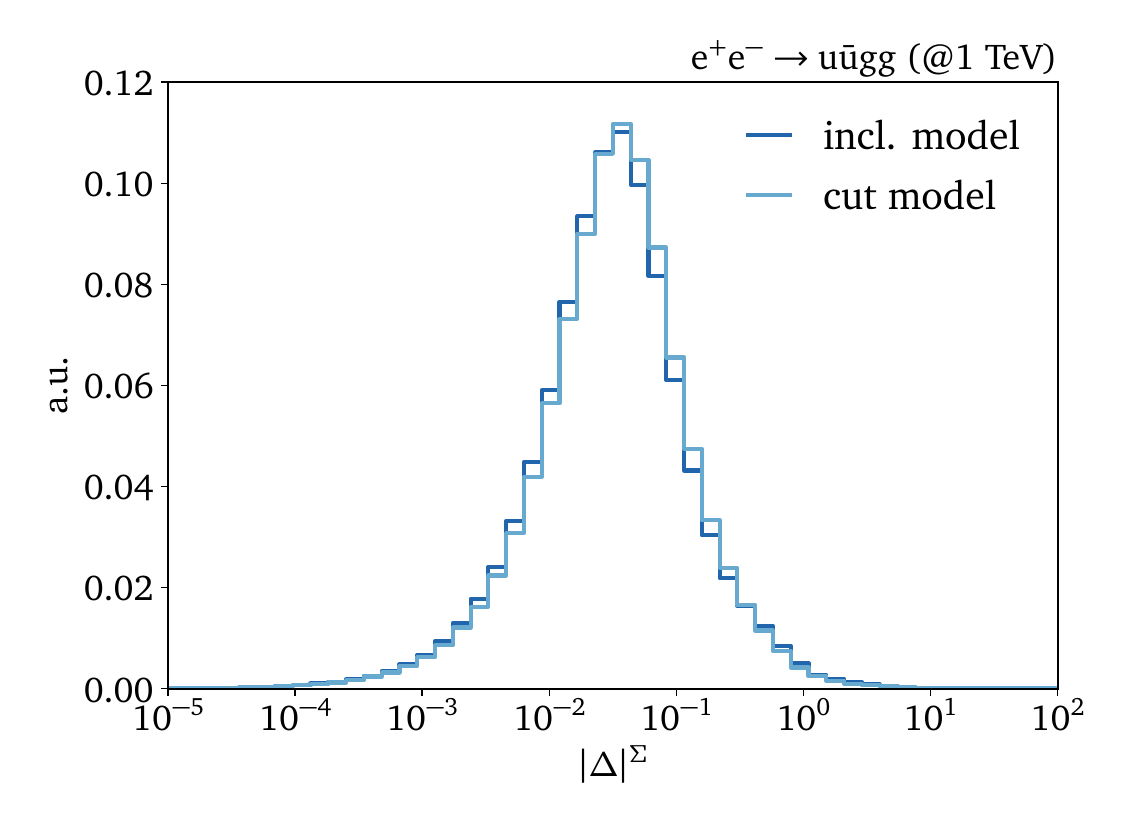} \\
\includegraphics[width=0.495\linewidth, page=2]{figs/np1body/new-cut-trainings/delta_cutdata_ee_3j.pdf}
\includegraphics[width=0.495\linewidth, page=2]{figs/np1body/new-cut-trainings/delta_cutdata_ee_4j.pdf}
\caption{Relative accuracies for the real-emission amplitudes (upper) and pull distributions (lower) for the NLO corrections to 3-jet (left) and 4-jet (right) production. We compare training on the full phase space (incl.) and outside the subtraction region (cut), both evaluated outside the subtraction region.}
\label{fig:target_and_benchmark_sigma_large_ds}
\end{figure} 
%-------------------------------

We only train on points with $\Sigma_{ij}\neq 0$.
This discards the soft-quark regions of $\Pep \Pem \to \Pu \Pubar \;\Pq \Pqbar$, which have no soft singularity and do not contribute to the integral.
Since the remaining $\Sigma_{ij}$ are positive, we learn their standardized logarithm.
The network is again a simple MLP.
We provide the discrete FKS sector index through a learned embedding, implemented as a look-up table followed by a linear layer.
This way, only a small set of parameters is sector specific, while all other layers are shared.
Conditioning all layers on the sector index does not improve the accuracy significantly.
As we only use the real-emission surrogates outside the subtraction region, we also test whether the training data can be restricted to this region.
The subtraction region is controlled by a single threshold $\lambda$, which we optimize in Sec.~\ref{sec:optimizing_subtraction} to $\lambda=0.05$ for the 3-jet and $\lambda=0.01$ for the 4-jet process.
We compare a network trained on the full real-emission phase space with one trained only outside the subtraction region, and evaluate both outside the subtraction region.
Training on the full phase space leads to at most a marginal improvement, so restricting the training data is sufficient.

In Fig.~\ref{fig:target_and_benchmark_sigma}, we compare the learned real-emission amplitudes to the truth over more than 15 orders of magnitude.
The relative accuracies in Fig.~\ref{fig:target_and_benchmark_sigma_large_ds} are clearly worse than for the virtual surrogates in Fig.~\ref{fig:benchmark_nbody}, even though we increase the training dataset from around 100k to 870k phase-space points per FKS sector.
Without this increase, in particular the sectors 4--6 without soft and soft--collinear singularities are not learned correctly.
Outside the subtraction region, the median relative deviation is around 1\% for the 3-jet and 4\% for the 4-jet process, with a tail towards deviations of order one.
There are three reasons for the lower precision.
First, the real-emission phase space includes one more final-state particle.
Second, there is no Born amplitude to factor out and no Lorentz-invariant parametrization.
Third, the amplitudes span twice as many orders of magnitude.
Despite the lower precision, the learned uncertainties remain calibrated, as shown by the pulls in the lower panels.
Finally, the test points are not weighted with the NLO cross section, so the relevant test of the surrogate precision is the cross-section computation in Sec.~\ref{sec:perf}.

Surrogates are hard to use where the subtraction is active, because the learned $\Sigma_{ij,\theta}$ is multiplied by the divergent factors $1/\xi_i^2$ and $1/(1-y_{ij})$, see Eq.\eqref{eq:real_sub}.
For instance, in the soft limit the FKS phase-space factor contributes one power of $\xi_i$, and the subtracted real-emission integrand obtained from a surrogate behaves as
\begin{align}
\dd\Phi^{(ij)}_{n+1}\,\mata_{ij, \theta}^{\text{R}-\text{S}}\Big|_{\xi_i\to0}
&\sim \frac{\dd\xi_i}{\xi_i}\;\left[
\Sigma_{ij,\theta}(\Phi_n,\xi_i,y_{ij},\varphi_i)
-\left|\frac{\partial\Phi^{\text{soft}}_{n+1}}{\partial\Phi^{\text{hard}}_{n+1}}\right|\,\Sigma_{ij,\theta}(\Phi_n,0,y_{ij},\varphi_i) \right]\; .
\label{eq:diff_surr}
\end{align}
For exact amplitudes, the two terms in the bracket approach each other for $\xi_i\to0$, so their difference vanishes like $\xi_i$ and the integrand stays finite.
For surrogates, the two terms are evaluated at different kinematic configurations, which are far apart in the logarithmic input variables, so their errors are uncorrelated and do not cancel in the difference.
For a relative surrogate accuracy $\epsilon$, the relative error of the subtracted amplitude then grows like
\begin{align}
    \frac{\Delta \mata_{ij,\theta}^{\text{R}-\text{S}}}{\mata_{ij}^{\text{R}-\text{S}}} \sim \frac{\epsilon}{\xi_i} \; ,
\end{align}
which leads to a logarithmically divergent error in the integral.
The same argument applies to the collinear limit.
We therefore do not use the real-emission surrogate wherever a subtraction is active and evaluate the exact amplitudes instead.
The default \mg setup evaluates subtracted amplitudes over most of the real-emission phase space.
To still benefit from the surrogate, we reduce the subtraction region by changing the subtraction parameters in Eq.\eqref{eq:default_cutoffs}.
A more nuanced approach, based on the calibrated uncertainties, could use a dynamic choice between surrogates and exact amplitudes~\cite{Beccatini:2025tpk} or a dedicated training~\cite{Bahl:2026qaf}.
However, neither solves the fundamental problem of extremely sensitive cancellations, which would require an efficient learning of the difference between functions~\cite{Butter:2019eyo}.

\clearpage
%%%%%%%%%%%%%%%%%%%%%%%%%%%%%%%%%%%%%%%%%%%%%%%%%%%
\section{MadNIS@NLO sampling}
\label{sec:sampling}

The second ingredient to ultrafast NLO calculations is neural importance sampling. To extend \madnis to NLO, we adapt the multi-channel formalism to include the FKS partitioning of the real-emission phase space, as defined in Eq.\eqref{eq:fks_expand}.

%%%%%%%%%%%%%%%%%%%%%%%%%%%%%%%%%%%%%%%%%%%%%%%%%%%
\subsection{Phase-space mappings}

%%%%%%%%%%%%%%%%%%%%%%%%%%%%%%%%%%%%%%%%%%%%%%%%%%%
\subsubsection*{Born-like phase space}
The Born-level phase space is sampled using a multi-channel setup, where each channel corresponds to a single topology of the tree-level Feynman diagrams. Diagrams differing only by a permutation of the final-state particles are integrated together. The integrand is divided into $N_c$ channels using the single-diagram enhancement strategy~\cite{Maltoni:2002qb,Mattelaer:2021xdr,Heimel:2022wyj,Heimel:2023ngj},
\begin{align}
    \sigma_n = \int \dd \Phi_n\; f_n(\Phi_n)
    = \sum_k \int \dd \Phi_n\: \alpha_k(\Phi_n)\;f_n(\Phi_n) 
    \qquad \mwith \qquad \sum_k \alpha_k(\Phi_n) =1 \;.
    \label{eq:ps_born}
\end{align}
We define $\alpha_k$ as a product of the propagator denominators~\cite{Mattelaer:2021xdr},
\begin{align}
    \alpha_k(\Phi_n) \propto \prod_{\text{propagators }\ell}
    \frac{1}{\left(p_\ell^2 - m_\ell^2\right)^2 + m_\ell^2 \Gamma_\ell^2} \; ,
\end{align}
and use \madspace~\cite{Heimel:2026hgp} to implement analytic channel mappings between the unit hypercube and the physical phase space
\begin{align}
    x_\text{B} \; \xleftrightarrow[\text{each channel }k]{\quad\text{mapping for}\quad} \; \Phi^{(k)}_n \; ,
\end{align}
with associated normalized sampling density
\begin{align}
   J^k_{\text{B}}(\Phi^{(k)}_n)=\left\vert\frac{\partial x_\text{B}(\Phi^{(k)}_n)}{\partial \Phi^{(k)}_n}\right\vert
   \qquad \mwith \qquad 
   \int \dd \Phi^{(k)}_n\;J^k_{\text{B}}(\Phi^{(k)}_n)=1  \; .
   \label{eq:channel_densities}
\end{align}
This allows us to rewrite Eq.\eqref{eq:ps_born} as
\begin{align}
    \sigma_n  
    &= \sum_k \int \dd x_\text{B}\;\alpha_k(\Phi^{(k)}_{n}(x_\text{B})) \;  \frac{f_n(\Phi^{(k)}_{n}(x_\text{B}))}{J^k_{\text{B}}(\Phi^{(k)}_{n}(x_\text{B}))}\notag\\
    &\equiv \sum_k \int \dd x_\text{B}\;\alpha_k(\Phi^{(k)}_n(x_\text{B}))\; w^k_{n}(x_\text{B}) \; .
    \label{eq:sigma_n_unit_hypercube}
\end{align}

%%%%%%%%%%%%%%%%%%%%%%%%%%%%%%%%%%%%%%%%%%%%%%%%%%%
\subsubsection*{Real-emission phase space}

Next, we target the FKS-partitioned real-emission contribution from Eq.\eqref{eq:fks_expand},
\begin{align}
\sigma_{n+1}=\sum_{ij} \int \dd \Phi^{(ij)}_{n+1}\;f^{ij}_{n+1}(\Phi^{(ij)}_{n+1})\; .
\label{eq:real_emission_ps}
\end{align}
Analogously to the channel mappings, we introduce a mapping in each FKS sector
\begin{align}
(\Phi_n,\xi_i,y_{ij},\varphi_i)\equiv(\Phi_n,\Phi^{ij}_\text{rad})\; \xleftrightarrow[\text{for each }ij]{\quad\text{FKS mapping}\quad} \; \Phi^{(ij)}_{n+1}\;.
\end{align}
This allows us to parametrize the phase-space integral as
\begin{align}
\sigma_{n+1}&=\sum_{ij} \int \dd \Phi_n\; \dd\Phi^{ij}_\text{rad}\;J^{ij}_\text{FKS}\bigl(\Phi_n,\Phi^{ij}_\text{rad}\bigr)\: f^{ij}_{n+1}\!\left(\Phi^{(ij)}_{n+1}(\Phi_n,\Phi^{ij}_\text{rad})\right) 
\notag\\
&\equiv\sum_{ij} \int \dd \Phi_n\; \dd\Phi^{ij}_\text{rad}\: h^{ij}_{n+1}(\Phi_n,\Phi^{ij}_\text{rad})\;,
\label{eq:real_emission_ps_rad}
\end{align}
with
\begin{align}
\dd \Phi^{(ij)}_{n+1}
= J^{ij}_\text{FKS}(\Phi_n,\Phi^{ij}_\text{rad})\;\dd \Phi_n\;
\dd\Phi^{ij}_\text{rad}
\qquad \mand \qquad
\dd\Phi^{ij}_\text{rad}\equiv
\dd \xi_i\,\dd y_{ij}\,\dd \varphi_{i}\; .
\label{eq:fks_ps_factorization_def}
\end{align}
The Jacobian $J^{ij}_\text{FKS}$ accounts for the map from the Born-like momenta $\Phi_n$ and the radiation variables to the $(n+1)$-body phase space $\Phi_{n+1}$.
To compute it, we consider a generic FKS splitting $\pmother \to \psister + \pdaughter$, with relevant momenta~\cite{Frixione:2007vw}
\begin{align}
    \text{Mother (emitting) parton: } &\pmother \notag \\
    \text{Sister (after emitting) parton: } &\psister \notag \\
    \text{Daughter (emitted) parton: } &\pdaughter \;.
\end{align}
We define the center-of-mass momentum and the recoil mass as
\begin{align}
    q=\sum_{k=1}^n p_k
    \qquad \mwith \qquad q^2=(q^0)^2=s
    \qquad \mand \qquad M_{j,\text{rec}}^2 = (q-p_j)^2 \;.
\end{align}
Using the definition of the radiation variables in Eq.\eqref{eq:rad_variables}, we immediately obtain
\begin{align}
    \tilde p^0_i = |\mathbf{\tilde p}_i| = \xi_i \frac{\sqrt{s}}{2} \; .
    \label{eq:p_i_fks}
\end{align}
Energy-momentum conservation then gives
\begin{align}
|\mathbf{\tilde p}_j|
=
\frac{s-M_{j,\text{rec}}^2-\xi_i\,s}{2\sqrt{s}-\xi_i(1-y_{ij})\sqrt{s}}
\qquad \mand \qquad
\tilde p^0_j = |\mathbf{\tilde p}_j|\;.
\end{align}
Next, we choose their directions such that $\mathbf{\tilde{p}}_{j}+\mathbf{\tilde{p}}_{i} \parallel \mathbf{p}_j$ and the azimuthal angle of $\mathbf{\tilde{p}}_{i}$ around the axis $\mathbf{\tilde{p}}_{j}+\mathbf{\tilde{p}}_{i}$ is $\varphi_i$.
Together with $y_{ij}$, this fully determines $\psister$ and $\pdaughter$, but the set $(p_1,\ldots,\psister,\ldots,p_{n},\pdaughter)$ does not satisfy 4-momentum conservation.
We therefore define the recoil momentum,
\begin{align}
    k_{ij,\text{rec}} = q - (\psister + \pdaughter) 
    \qquad \mand \qquad 
    \mathbf{k}_{ij,\text{rec}} = - \mathbf{\tilde{p}}_{j} - \mathbf{\tilde{p}}_{i}\;,
\end{align}
and construct a boost $\Lambda_{\beta_{ij}}$ along $\mathbf{k}_{ij,\text{rec}}$, with boost parameter $\beta_{ij}$, which maps it onto the Born recoil momentum $q-p_j$, so that
\begin{align}
    (q-\Lambda_{\beta_{ij}}\,k_{ij,\text{rec}})^2=0
    \qquad \mwith \qquad
    \beta_{ij} = \frac{s-(k^{0}_{ij,\text{rec}} + |\mathbf{k}_{ij,\text{rec}}|)^{2}}{s+(k^{0}_{ij,\text{rec}} + |\mathbf{k}_{ij,\text{rec}}|)^{2}} \; .
\end{align}
This allows us to obtain the remaining momenta through the inverse boost of the Born-like momenta,
\begin{align}
    \tilde{p}_k = \Lambda^{-1}_{\beta_{ij}}\,p_k
    \qquad \mfor \qquad
    k \neq i,j \; .
\end{align}
The FKS Jacobian is then given by
\begin{align}
J^{ij}_\text{FKS}(\Phi_n,\Phi^{ij}_\text{rad})
=
\xi_i\frac{s}{(4\pi)^3}\;
\frac{|\mathbf{\tilde p}_j|^2}{|\mathbf p_j|}
\left(
|\mathbf{\tilde p}_j| - \frac{(\tilde p_j+\tilde p_i)^2}{2\sqrt{s}}
\right)^{-1}\;.
\label{eq:jac_fks}
\end{align}
After decomposing the real-emission phase space into Born-like and radiation phase spaces, we again impose the multi-channel splitting from Eq.\eqref{eq:ps_born} and rewrite Eq.\eqref{eq:real_emission_ps_rad} as
\begin{align}
\sigma_{n+1}=\sum_k \sum_{ij} \int \dd \Phi_n\;\dd\Phi^{ij}_\text{rad}\;\alpha_k(\Phi_n)\: h^{ij}_{n+1}(\Phi_n,\Phi^{ij}_\text{rad}) \;.
\label{eq:fks_and_mc}
\end{align}
This allows us to introduce channel mappings from an enlarged unit hypercube and hence the complete mapping
\begin{align}
(x_\text{B}, x_\text{rad})\;
\xleftrightarrow[\text{each channel }k]{\quad\text{mapping for}\quad}
\;(\Phi^{(k)}_n,\Phi^{ij}_\text{rad})\;\xleftrightarrow[\text{for each FKS sector }ij]{\quad\,\text{mapping for}\,\quad}\; \Phi^{(k,ij)}_{n+1} \; .
\label{eq:def_chan_fks}
\end{align}
We parametrize the radiation phase space by three independent unit hypercube
variables $x_\text{rad}=(x_\xi,x_y,x_\varphi)\in[0,1]^3$ with the discrete FKS pair $ij\in{\cal P}_\text{FKS}$, which we sample uniformly before the \madnis training. 
Similarly to what is currently done in \mg, we then define
\begin{align}
y_{ij}(x_y) &= 1-2\,x_y^2  \notag \\
\varphi_i(x_\varphi) &= 2\pi\,x_\varphi \notag\\
\xi_i(x_\xi) &= \xi_{j,\max}\,x_\xi^2  \qqquad \mwith \qquad
\xi_{j,\max} = (s-M_{j,\text{rec}}^{2})/s \;,
\end{align}
with the corresponding Jacobian
\begin{align}
J^{ij}_\text{rad}(\Phi^{ij}_{\text{rad}})
=\frac{1}{16\pi}\;
\frac{1}{\sqrt{\xi_{j,\max}\,\xi_i}}\;
\sqrt{\frac{2}{1-y_{ij}}}\;.
\label{eq:rad_map}
\end{align}
The quadratic remappings $x_\xi \mapsto \xi_i \propto x_\xi^2$ and $x_y \mapsto y_{ij}=1-2x_y^2$ absorb the remaining integrable soft and collinear singularities of the radiation phase space analytically into the phase-space measure, since they would otherwise lead to a large variance.

For the Born-like part of the real-emission phase space, we use the same multi-channel mappings as in the Born contribution, so for each channel $k$ we map $x_\text{B}$ to the Born kinematics. Combining this with the radiation map above, we find for Eq.\eqref{eq:fks_and_mc} 
\begin{align}
\sigma_{n+1}
&=\sum_k \sum_{ij}
\int \dd x_\text{B}\,\dd x_\text{rad}\;
\alpha_k\bigl(\Phi_n^{(k)}(x_\text{B})\bigr)\;
\frac{h^{ij}_{n+1}\Bigl(\Phi_n^{(k)}(x_\text{B}),\Phi_\text{rad}^{ij}(x_\text{rad})\Bigr)}{J^k_\text{B}(\Phi_n^{(k)}(x_\text{B}))\;J^{ij}_\text{rad}\bigl(\Phi_\text{rad}^{ij}(x_\text{rad})\bigr)}\notag\\
&\equiv\sum_k \sum_{ij}
\int \dd x_\text{B}\,\dd x_\text{rad}\;
\alpha_k\bigl(\Phi_n^{(k)}(x_\text{B})\bigr)\;
w^{k,ij}_{n+1}(x_\text{B},x_\text{rad})
\;.
\label{eq:sigma_nplus_unit_hypercube}
\end{align}
For fixed-order NLO computations in \mg, both Born-like and real-emission kinematics stem from the same Born-like momenta and are sampled together,
\begin{align}
\sigma_\text{NLO} &= \sum_k \sum_{ij} \int \dd x_\text{B}\,\dd x_\text{rad}\; \alpha_k\bigl(\Phi_n^{(k)}(x_\text{B})\bigr)\; \left[
    \frac{w^k_n(x_\text{B})}{n_\text{FKS}} + w^{k,ij}_{n+1}(x_\text{B},x_\text{rad})
\right] \notag \\
&\equiv 
\sum_k \sum_{ij} \int \dd x_\text{B}\, \dd x_\text{rad} \; \alpha_k\bigl(\Phi_n^{(k)}(x_\text{B})\bigr)\; w_{\text{NLO}}^{k,ij}(x_\text{B},x_\text{rad})\; .
\label{eq:comb_ps_integral}
\end{align}
where $n_\text{FKS}$ denotes the number of FKS sectors.

%%%%%%%%%%%%%%%%%%%%%%%%%%%%%%%%%%%%%%%%%%%%%%%%%%%
\subsection{Neural importance sampling}

%------------------------------------------
\begin{figure}[t]
    \centering
    \input{figs/tikz/fks}
    \caption{Illustration of the sampling and evaluation of phase-space points at NLO for a given integration channel $k$. The boxes with a violet border represent building blocks that are augmented with ML using either MadNIS (green boxes) or amplitude surrogates (red boxes).}
    \label{fig:fks}
\end{figure}
%------------------------------------------

Finally, we employ \madnis to map latent variables $(z_\text{B}, z_\text{rad})$ onto the unit hypercube and smooth out the integrand in Eq.\eqref{eq:comb_ps_integral},
\begin{align}
(z_\text{B}, z_\text{rad})
\xleftrightarrow[k,\;ij]{\;\quad\text{MadNIS}\quad\;}
(x_\text{B}, x_\text{rad})
\xleftrightarrow[k,\;ij]{\quad\text{chan. mapping}\quad}
(\Phi^{(k)}_n,\Phi^{ij}_\text{rad})
\xleftrightarrow[ij]{\quad\text{FKS mapping}\quad}
\Phi^{(k,ij)}_{n+1} \; .
\label{eq:def_inn_and_rest}
\end{align}
As illustrated in Fig.~\ref{fig:fks}, we use an independent normalizing flow for each channel $k$, each with parameters $\theta_k$, and each of them conditioned on the FKS sector $ij$.
Channels related by a symmetry share a single flow.
We start by sampling the discrete FKS index $ij$, using a vector of learned log-probabilities.
While the optimal probabilities, assuming perfectly trained normalizing flows, could in principle be obtained directly by computing the absolute per-sector integrals, we optimize them along with the normalizing flow parameters through gradient descent. This part of the sampling is thus ML-augmented only in the sense that it is learned jointly with the flows. The normalized probability $p_{\theta_k}(ij)$ is obtained from a softmax function. Then we sample the continuous $x_\text{B}$ and $x_\text{rad}$ jointly using a normalizing flow conditioned on a one-hot encoding of $ij$
\begin{align}
    g_{\theta_k}(x_\text{B}, x_\text{rad}, ij) =
    q_{\theta_k}(x_\text{B}, x_\text{rad} | ij) \; p_{\theta_k}(ij) \;.
\end{align}
The multi-channel NLO-\madnis integral then becomes
\begin{align}
   \sigma_\text{NLO} = \sum_k \left\langle \frac{\alpha_{\psi,k}\bigl(\Phi_n^{(k)}(x_\text{B})\bigr)\; w_{\text{NLO}}^{k,ij}(x_\text{B},x_\text{rad})}{q_{\theta_k}(x_\text{B}, x_\text{rad} | ij) \: p_{\theta_k}(ij)} \right\rangle_{\hspace{-1mm}\substack{ij \sim p_{\theta_k}(ij)\qqquad \\ (x_\text{B}, x_\text{rad}) \sim q_{\theta_k}(x_\text{B}, x_\text{rad} | ij)}} \;,
\end{align}
where $\theta_k$ are the parameters of the normalizing flow for channel $k$, and $\alpha_{\psi,k}$ are learned channel weights with network parameters $\psi$, which refine the analytic $\alpha_k$ from Eq.\eqref{eq:ps_born}~\cite{Heimel:2022wyj,Heimel:2023ngj}.
This estimate is unbiased for any sampling density that covers the support of the integrand, because neural importance sampling only affects the variance, and not the expected value of the estimated Monte Carlo integral.
A bias only appears when we replace the exact integrand by surrogates, and we show in Sec.~\ref{sec:perf} that it stays below the Monte Carlo uncertainty for the statistics used here.

We perform a standard online \madnis training, following Ref.~\cite{Heimel:2023ngj}.
After a short \vegas pretraining, the (clipped) multi-channel stratified variance loss is minimized jointly over the flow parameters $\theta_k$ and the channel weight network parameters $\psi$.
The training hyperparameters are listed in Tab.~\ref{tab:madnis_hyperparams}.

At LO, the performance of a sampler is usually quantified by its unweighting efficiency.
Unweighting is not defined in a fixed-order NLO calculation, so we instead use the relative variance of the integral
\begin{align}
    \frac{\text{Var}(w_{\text{NLO}})}{\sigma^2_{\text{NLO}}} \; .
\end{align}
For $N$ samples, the relative integration error is
\begin{align}
    \varepsilon^2 = \frac{1}{N}\,\frac{\text{Var}(w_{\text{NLO}})}{\sigma^2_{\text{NLO}}} \; ,
    \label{eq:rel_error}
\end{align}
so the ratio of relative variances from two samplers gives the ratio of samples needed to reach a given precision, \ie the integration acceleration.
At LO, a smaller relative variance typically comes with a higher unweighting efficiency~\cite{Heimel:2023ngj}.
Unweighted NLO event generation requires matching to a parton shower and is beyond the scope of this study.

%\clearpage
%%%%%%%%%%%%%%%%%%%%%%%%%%%%%%%%%%%%%%%
\section{Performance}
\label{sec:perf}

After testing the surrogates independently in Sec.~\ref{sec:surrogates} and introducing the FKS-conditioned \madnis in Sec.~\ref{sec:sampling}, we now turn to the performance for NLO predictions of 3-jet and 4-jet production in Eq.\eqref{eq:def_procs_lo}.
We use the virtual surrogate throughout, but follow the conservative approach of only using the real-emission surrogate where no subtraction is active.
We therefore first optimize the size of the subtraction region and then show the precision and acceleration of \madnis at NLO.

%%%%%%%%%%%%%%%%%%%%%%%%%%%%%%%%%%%
\subsection{Optimized subtraction threshold} 
\label{sec:optimizing_subtraction}

When employing real-emission surrogates in a subtraction scheme, we face two challenges:
\begin{enumerate}[label=(\roman*)]
    \item Even per-mille surrogate accuracy for $\Sigma_{ij,\theta}$ is insufficient to reproduce the delicate cancellations required in the soft and collinear regions, as discussed in Sec.~\ref{sec:surrogates}.
    \item Learning the subtracted combination $\mata^{\text{R-S}}_{ij}$ directly would avoid these cancellations.
    However, through the $\Theta$ functions in Eq.\eqref{eq:real_sub} such a surrogate depends on $\xi_\text{cut}$ and $\delta$, so it has to be retrained for every threshold choice and is discontinuous at the threshold boundaries.
    Moreover, the real-emission term and its counterterms are evaluated at different kinematic configurations, which pass the cuts independently and enter different bins of kinematic distributions.
    In this work, we therefore restrict ourselves to surrogates for $\Sigma_{ij}$ away from the divergent limits.
\end{enumerate}
In the standard \mg setup, the subtraction regions defined by Eq.\eqref{eq:default_cutoffs} cover a large fraction of the real-emission phase space. While such an extended subtraction support is not strictly required for convergence, it reduces the fraction of negative integrands, \ie $w^{k,ij}_\text{NLO}<0$, and thereby lowers the Monte Carlo variance. 

%-------------------------------
\begin{figure}[b!]
\includegraphics[width=0.495\linewidth, page=6]{figs/madnis/sweep_threshold.pdf}
\includegraphics[width=0.495\linewidth, page=7]{figs/madnis/sweep_threshold.pdf}\\
\includegraphics[width=0.495\linewidth, page=2]{figs/madnis/sweep_threshold.pdf}
\includegraphics[width=0.495\linewidth, page=8]{figs/madnis/sweep_threshold.pdf}
\caption{Upper: relative variance as a function of the soft and collinear subtraction thresholds using the exact matrix element (left) and the real-emission surrogate outside the subtraction region (right). Lower: cross section computed using the real-emission surrogate (left) and fraction of surrogate evaluations (right). All panels show 3-jet production.}
\label{fig:threshold_fks1}
\end{figure} 
%------------------------------

However, this choice is not optimal when using a real-emission surrogate $\Sigma_{ij,\theta}$ that can only be efficiently employed in the non-subtracted phase-space regions. We therefore re-optimize the subtraction thresholds by balancing the integrand variance against the potential speed gains from the surrogate in three ways:
\begin{alignat}{3}
\text{collinear threshold}&
& \qquad \delta&=\lambda 
& \qquad \xi_{\text{cut}}&=0.5 \notag \\
\text{soft threshold}&
& \delta&=1.0
&  \xi_{\text{cut}}&=\lambda\notag \\
\text{combined thresholds}&
& \delta&=\lambda
& \xi_{\text{cut}}&=\lambda \qqquad \mwith \qquad \lambda\in[10^{-4}, 1]\;.
\label{eq:cut_paths}
\end{alignat}
In the upper panels of Fig.~\ref{fig:threshold_fks1}, we show the relative variance for different values of $\lambda$, which increases with less subtraction and explains why the standard \mg implementation chooses a subtraction over most of phase space.
This holds for the exact matrix elements (left) as well as for the surrogate (right) outside the subtraction region.

In the bottom left panel of Fig.~\ref{fig:threshold_fks1} we show the integrated cross section using the real surrogate. It is stable over a wide range of threshold values. However, in the bottom right panel we show the benefit of smaller threshold values as the fraction $f_\lambda$ of surrogate calls over the real-emission phase space increases, accelerating the numerical evaluation. In the following, we vary the soft and collinear threshold simultaneously, $\lambda=\delta = \xi_{\text{cut}}$, leaving a more detailed optimization to the final implementation.

%-------------------------------
\begin{figure}[t]
\centering
    \includegraphics[width=0.5\linewidth, page=1]{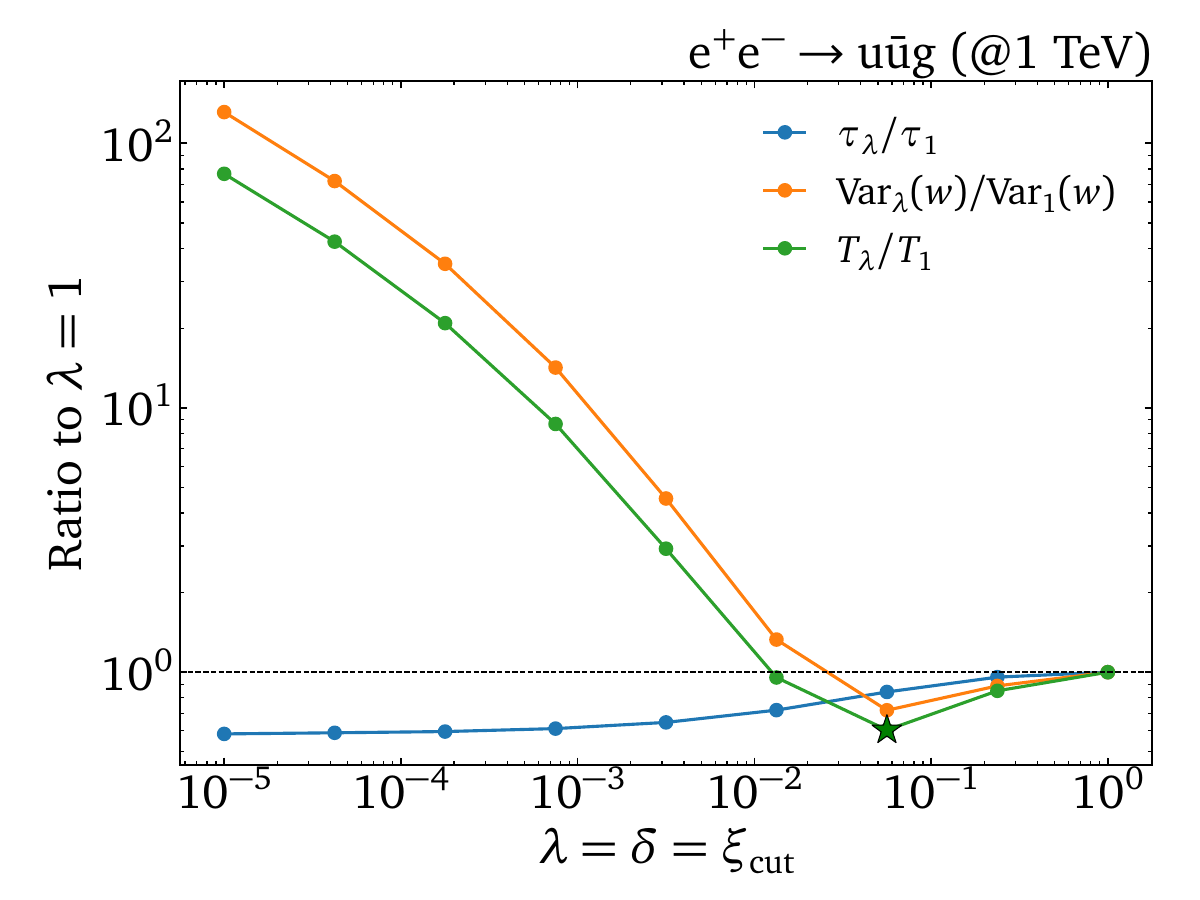}
    \caption{Average evaluation time per phase-space point, relative variance, and total evaluation time as a function of the subtraction threshold $\lambda$ for 3-jet production. All curves are normalized to the reference at $\lambda=1$.}
    \label{fig:FOM}
\end{figure} 
%------------------------------

A smaller $\lambda$ reduces the evaluation time per phase-space point, but it increases the variance and therefore the number of points needed for a given relative precision $\varepsilon$.
From Eq.~\eqref{eq:rel_error}, this number is
\begin{align}
    N_\lambda(\varepsilon) = \frac{\text{Var}_\lambda(w)}{\varepsilon^2\sigma^2_\text{NLO}} \; ,
\end{align}
where $\text{Var}_\lambda(w)$ is the variance of $w_\text{NLO}$ for a given $\lambda$.
We always use the virtual surrogate $R^{\mathrm{V/B}}_\theta$.
The average integrand evaluation time $\tau_\lambda$ then depends on the fraction $f_\lambda$ of real-emission points evaluated with $\Sigma_{ij,\theta}$, and we minimize the total evaluation time
\begin{align}
     T_\lambda(\varepsilon) \equiv N_\lambda (\varepsilon)\,\tau_\lambda
    = \frac{\text{Var}_\lambda(w)}{\varepsilon^2\sigma_\text{NLO}^2}
    \;
    \left[ f_\lambda\,\tau_{R^{\mathrm{V/B}}_\theta+\Sigma_{ij,\theta}}
    + (1-f_\lambda)\,\tau_{R^{\mathrm{V/B}}_\theta + \Sigma_{ij,\text{exact}}} \right] \; .
\end{align}
Figure~\ref{fig:FOM} shows $\tau_\lambda$, $\text{Var}_\lambda(w)$, and $T_\lambda$ for the 3-jet process, normalized to $\lambda=1$.
The minimum defines our settings, where the 4-jet optimization works analogously and is not shown,
\begin{alignat}{7}
 \text{3-jet:} &&\qqqquad 
 \lambda &\approx 0.05 
 \qquad \text{with} \qquad 
 f_\lambda \approx 40\%  \notag \\
 \text{4-jet:} &&\qqqquad 
 \lambda &\approx 0.01 
 \qquad \text{with} \qquad 
 f_\lambda \approx 65\%  \; .
\end{alignat}
%

%%%%%%%%%%%%%%%%%%%%%%%%%%%%%%%%%%%%%%%
\subsection{Precision and acceleration}

\begin{figure}[tp]
    \includegraphics[width=0.495\linewidth, page=1]{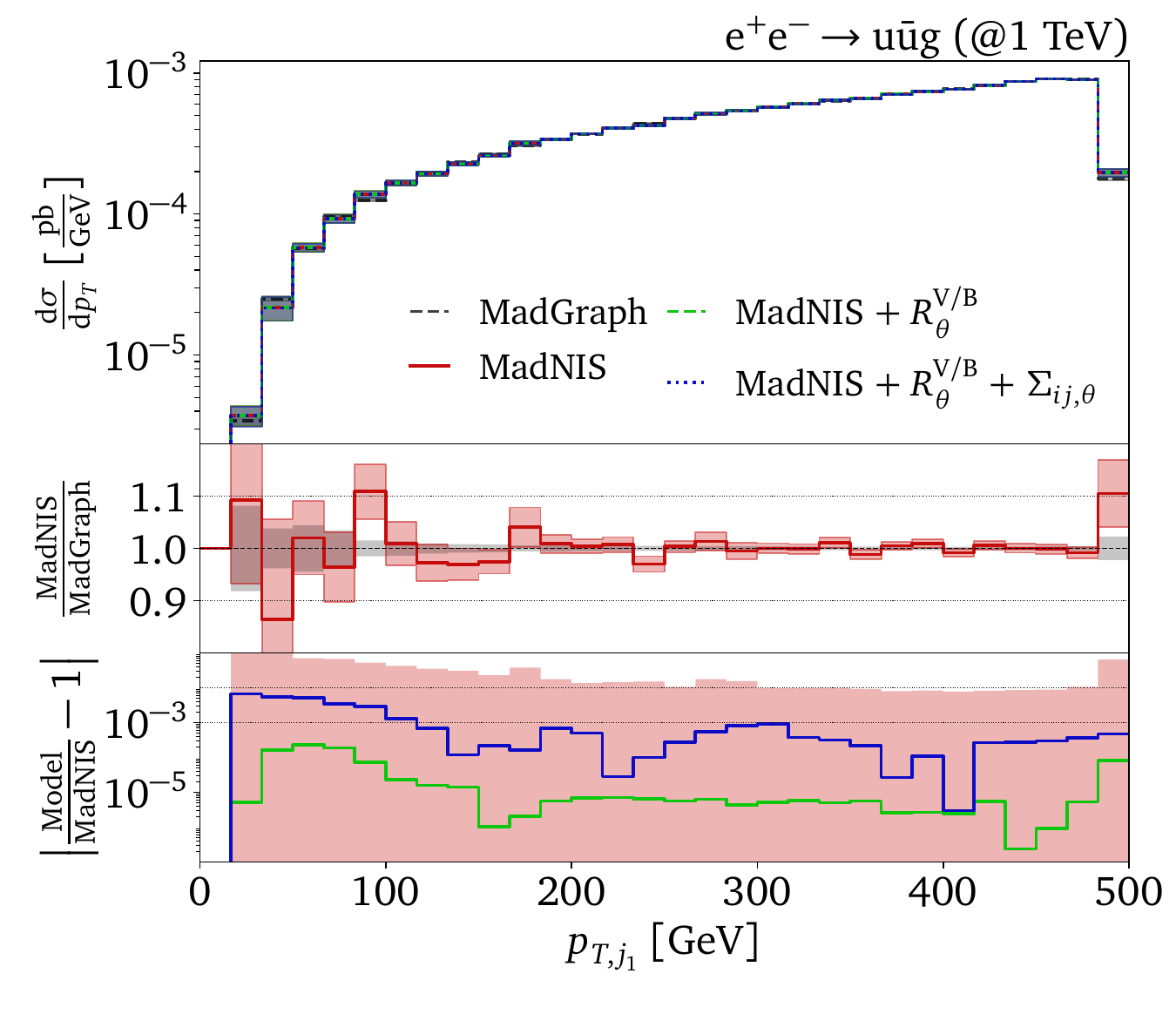}
    \includegraphics[width=0.495\linewidth, page=4]{figs/madnis/ee_3j/observables.pdf} \\
    \includegraphics[width=0.495\linewidth, page=18]{figs/madnis/ee_3j/observables.pdf}
    \includegraphics[width=0.495\linewidth, page=11]{figs/madnis/ee_3j/observables.pdf} \\
    \includegraphics[width=0.495\linewidth, page=28]{figs/madnis/ee_3j/observables.pdf}
    \includegraphics[width=0.495\linewidth, page=23]{figs/madnis/ee_3j/observables.pdf}
    \caption{Distributions of selected observables from 100M weighted \madnis events for $\Pep \Pem \to \Pu \Pubar \Pg$. We compare \mg with \vegas (black dashed) to \madnis with exact amplitudes (red solid), with the virtual-to-Born ratio surrogate (green dashed), and with both virtual-to-Born and real-emission surrogates (blue dotted). The red error bands of the lowest panel are the same as in the middle panel, and the dashed horizontal lines correspond to $0.1\%$ and $1\%$ level. The green line correctly does not appear for observables including the extra jet, since it involves only real-emission amplitudes.}
    \label{fig:madnis_observables-3j}
\end{figure}

\begin{figure}[tp]
    \includegraphics[width=0.495\linewidth, page=1]{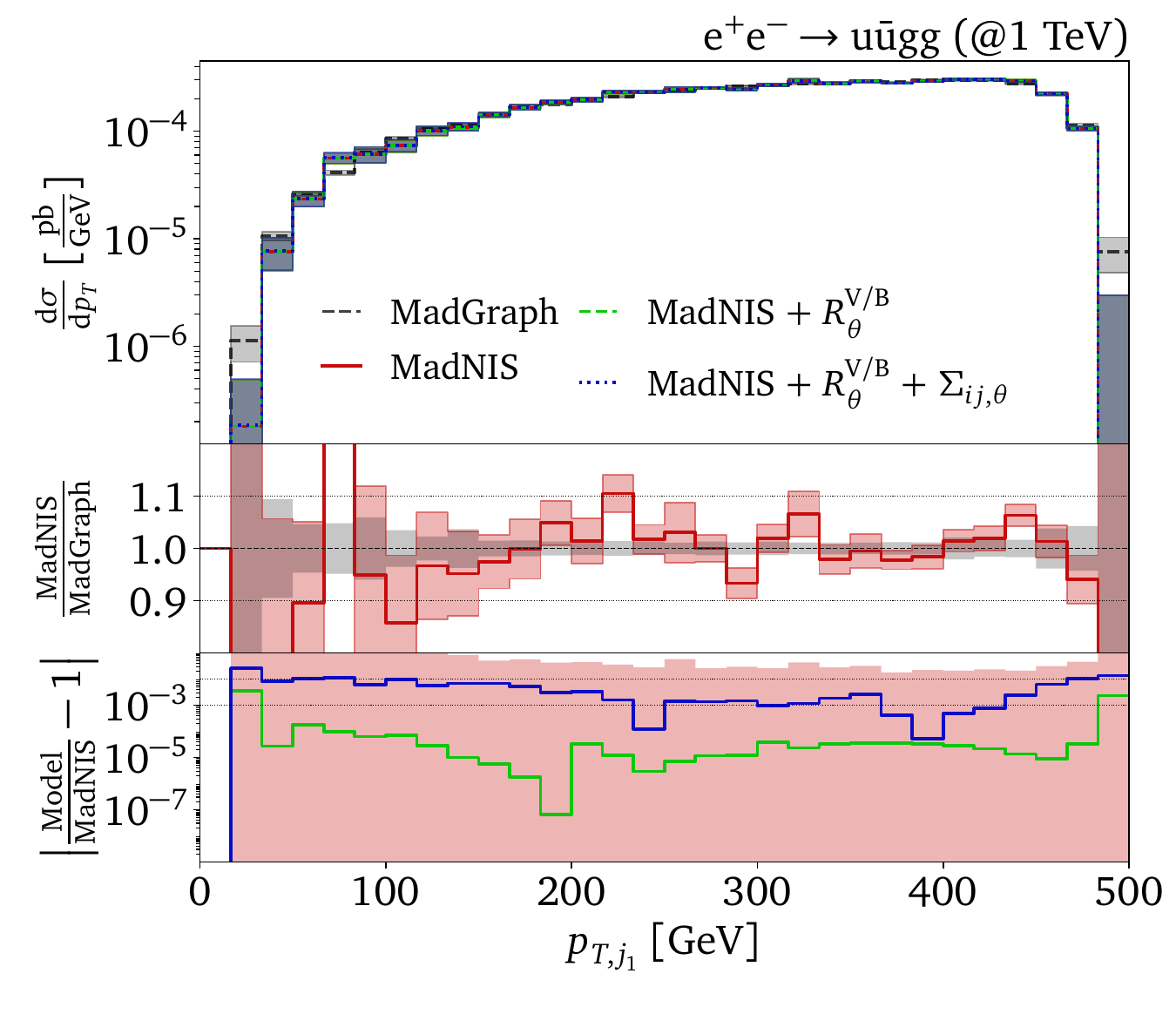}
    \includegraphics[width=0.495\linewidth, page=2]{figs/madnis/ee_4j/observables.pdf}\\
    \includegraphics[width=0.495\linewidth, page=9]{figs/madnis/ee_4j/observables.pdf}
    \includegraphics[width=0.495\linewidth, page=8]{figs/madnis/ee_4j/observables.pdf}\\    
    \includegraphics[width=0.495\linewidth, page=11]{figs/madnis/ee_4j/observables.pdf}
    \includegraphics[width=0.495\linewidth, page=10]{figs/madnis/ee_4j/observables.pdf}
    \caption{Distributions of selected observables from 500M weighted \madnis events for $\Pep \Pem \to \Pu \Pubar \Pg \Pg$. We compare \mg with \vegas (black dashed) to \madnis with exact amplitudes (red solid), with the virtual-to-Born ratio surrogate (green dashed), and with both virtual-to-Born and real-emission surrogates (blue dotted). The red error bands of the lowest panel are the same as in the middle panel, and the dashed horizontal lines correspond to $0.1\%$ and $1\%$ level. The green line correctly does not appear for observables including the extra jet, since it involves only real-emission amplitudes.}
    \label{fig:madnis_observables-4j}
\end{figure}

To validate our combined \madnis and surrogate methodology, we first study weighted histograms of kinematic observables using \madnis, with three ways of evaluating the amplitudes
\begin{enumerate}
    \item exact amplitudes only,
    \item the virtual-to-Born surrogate $R^{\text{V/B}}_\theta$,
    \item both surrogates, $R^{\text{V/B}}_\theta$ and $\Sigma_{ij,\theta}$.
\end{enumerate}
Figures~\ref{fig:madnis_observables-3j} and~\ref{fig:madnis_observables-4j} show the results for the 3-jet and 4-jet processes.

%-------------------------------
%------------------------------

%-------------------------------
%------------------------------

The second panels show the bin-wise ratio between \madnis with exact amplitudes and \mg, which agree throughout phase space.
The third panels compare \madnis with surrogates to \madnis with exact amplitudes, which directly quantifies the bias from the surrogates.
In the bulk of the distributions, the deviations stay below one per mille for the virtual surrogate and at the level of a few per mille with both surrogates.
In the tails and in a few individual bins, the deviations with both surrogates reach the percent level, but they stay below the Monte Carlo uncertainty of the \madnis prediction, shown as a band in the third panels.
We also show this bias for the integrated cross sections in Tab.~\ref{tab:results}.
Since this bias does not decrease with the number of events, much larger event samples eventually require a correction with exact amplitudes, as discussed below Eq.\eqref{eq:speedup}.

%------------------------------
\begin{table}[htbp]
    \setlength{\tabcolsep}{10pt}
    \centering
    \begin{small}
    \resizebox{\textwidth}{!}{
        \begin{tabular}{c|cc|S[table-format=1.5(2)]S[table-format=3.0(2)]|S[table-format=1.5(2)]S[table-format=4.0(2)]}
            \toprule
            Sampling % \multirow{2}{*}{sampling}
            & \multicolumn{2}{c|}{Surrogates}
            & \multicolumn{2}{c|}{$ \Pep \Pem \to \Pu \Pubar \Pg$} 
            & \multicolumn{2}{c}{$ \Pep \Pem \to \Pu \Pubar \Pg \Pg$} \\
            mode
            & {$R_\theta^{\text{V}/\text{B}}$} & {$ \Sigma_{ij, \theta}$}
            & {$\sigma_{\text{NLO}}\ [\mathrm{pb}]$}
            & {$\text{Var}(w_{\text{NLO}})/\sigma_{\text{NLO}}^2$} 
            & {$\sigma_{\text{NLO}}\ [\mathrm{pb}]$}
            & {$\text{Var}(w_{\text{NLO}})/\sigma_{\text{NLO}}^2$} \\
            \cmidrule(lr){1-7}
            \vegas & \ding{55} & \ding{55} & 0.10750(34) & 100(14) & 0.08721(15) & 2900(500) \\
            \madnis & \ding{55} & \ding{55} & 0.10760(19) & 30.4(2.6)  & 0.08729(15) & 720(90) \\
            \madnis & \ding{51} & \ding{55} &  0.10759(18) & 28.8(2.9)  &  0.08711(18) & 870(160)  \\
            \madnis & \ding{51} & \ding{51} &  0.10765(18)  & 27.4(1.1) &  0.08738(15) & 730(50) \\
            \bottomrule
        \end{tabular}
        }
    \end{small}
    \caption{Cross section and relative variance for each sampling and surrogate setup. Each value gives the mean over 5 runs (4-jet VEGAS uses 10 runs). In parentheses, we give the standard error of the mean for $\sigma_\text{NLO}$ and the standard deviation over the runs for the relative variance, \eg $30.4(26)=30.4\pm2.6$.}
    \label{tab:results}
\end{table}
%------------------------------
As a quantitative diagnostic of the integration performance of the combination of \madnis with fast ML-surrogates, we compare our three \madnis setups with standard \vegas adaptive sampling in Tab.~\ref{tab:results}. First, we determine the \vegas settings with a grid search for each process, minimizing the standard deviation of the integral, which gives the hyperparameters shown in Tab.~\ref{tab:vegas_hyperparams}. We then tune the number of phase-space points needed by \vegas for approximately 1\% precision. Next, we run \madnis with a short \vegas pretraining and the same number of points, leading to the hyperparameters shown in Tab.~\ref{tab:madnis_hyperparams}. We perform five runs for each setup and report the mean with uncertainties as specified in Tab.~\ref{tab:results}.
For the 4-jet \vegas baseline, we use $10^8$ points per run and ten runs.
The compatible relative variances of all \madnis runs confirm that the integration using surrogates is stable. While we observe agreement within uncertainties in the integrated cross sections, the relative variances indicate that we need three to four times more phase-space points with \vegas to reach \madnis precision, without and with surrogates.

%-------------------------------
\begin{figure}[htbp]
    \centering
    \includegraphics[width=0.495\linewidth, page=1]{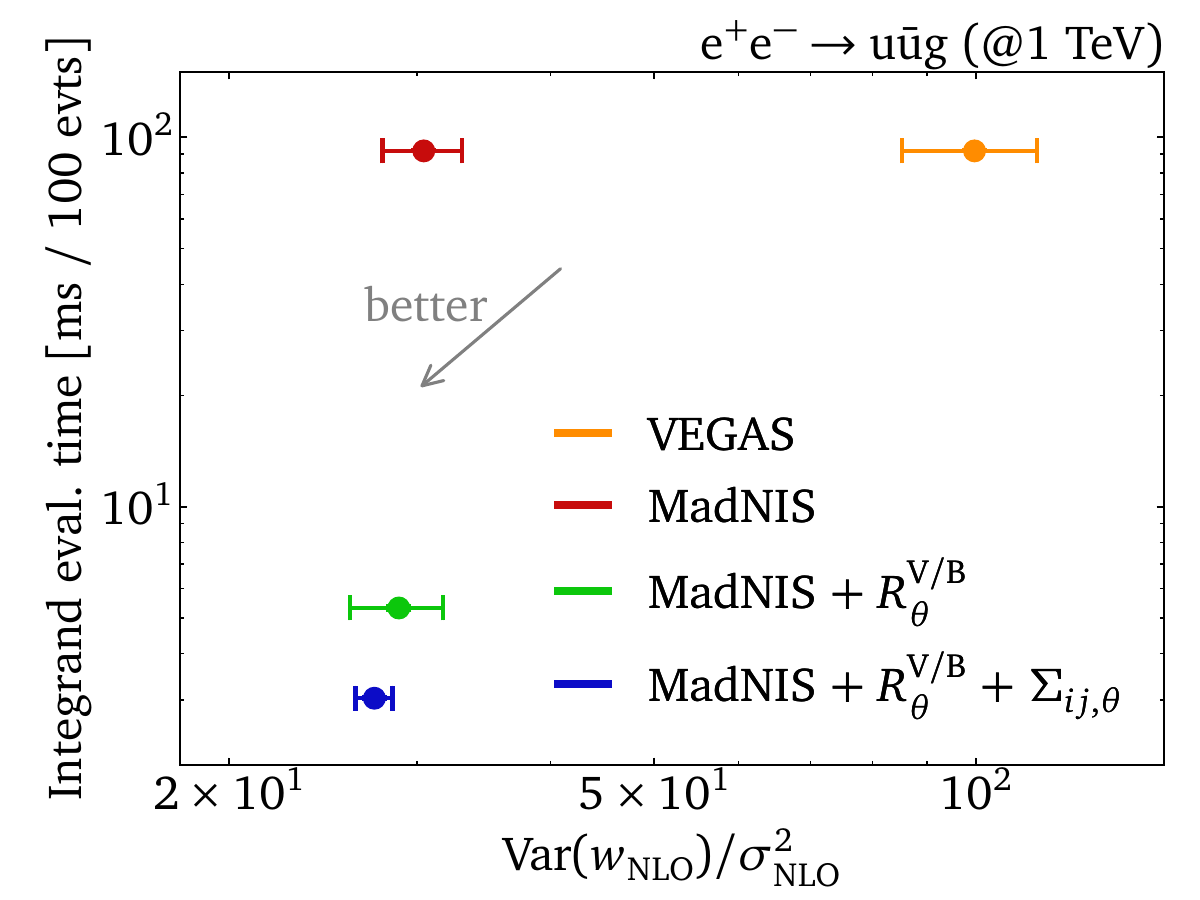}
    \includegraphics[width=0.495\linewidth, page=1]{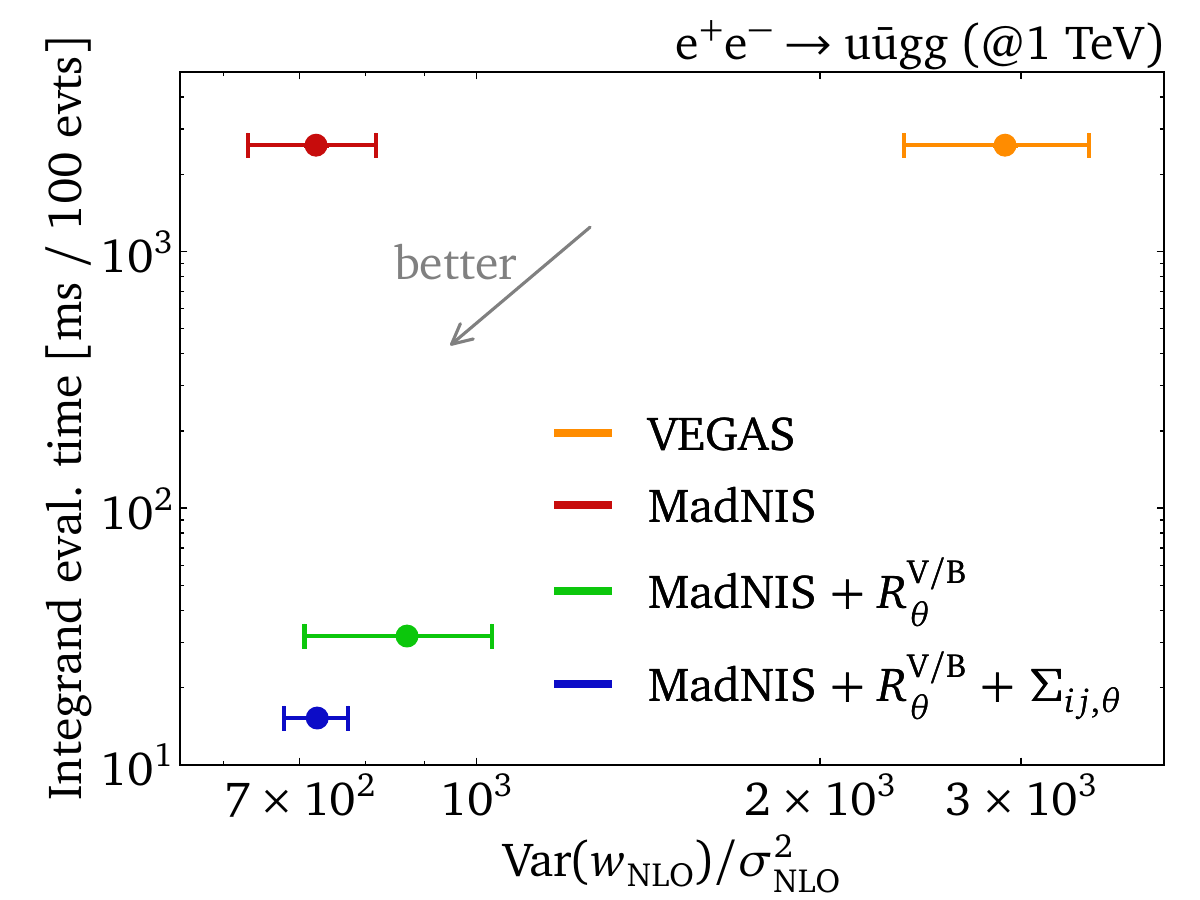}
    \caption{Average integrand evaluation time versus relative variance of the integral for 3-jet (left) and 4-jet (right) production. Lower left is better.}
    \label{fig:timing_vs_rv}
\end{figure} 
%------------------------------

The acceleration through ML-surrogates is shown in Fig.~\ref{fig:timing_vs_rv}, where we show the average integrand evaluation time versus relative variance of the integral. We report the mean and the standard deviation of five evaluations of 100 events on a single-core CPU. As expected, we find the same evaluation times for \madnis with exact amplitudes and the \vegas benchmark, but with a three times smaller relative variance.
These timings refer to the evaluation of the NLO integrand.
In our current implementation, the flow sampling, phase-space mapping, and cuts are not optimized and cost more than the surrogate-based integrand evaluation, so the end-to-end acceleration is smaller than the one quoted below.
However, since these steps do not depend on the matrix elements and can be compiled and parallelized, we use the integrand evaluation time as the relevant measure.

Switching to surrogates, we find an additional significant acceleration. The main driver of this acceleration is the virtual surrogate, which is a factor of 70 faster than the exact virtual amplitude for the 3-jet case and a factor of 600 for the 4-jet case. The combined acceleration of our 3-jet and 4-jet NLO predictions relative to \vegas sampling with exact amplitudes, with a surrogate-induced shift of the cross section below the Monte Carlo uncertainty of our runs, comes to
\begin{alignat}{7} 
\label{eq:speedup}
 \text{3-jet:} &&\qqqquad \qquad 
 \frac{T_{\madnis + R_\theta^{\text{V/B}}}}{T_\vegas} &\approx \frac{1}{60}
 &\qqqquad 
 \frac{T_{\madnis + R_\theta^{\text{V/B}} + \Sigma_{ij,\theta}}}{T_\vegas} &\approx \frac{1}{110} \notag \\
 \text{4-jet:} &&\qqqquad \qquad 
 \frac{T_{\madnis + R_\theta^{\text{V/B}}}}{T_\vegas} &\approx \frac{1}{230}
 &\qqquad 
 \frac{T_{\madnis + R_\theta^{\text{V/B}} + \Sigma_{ij,\theta}}}{T_\vegas} &\approx \frac{1}{680} \; .
\end{alignat}
Our \vegas baseline is comparable to \mg with the difference that the one-loop amplitudes are evaluated numerically at every phase-space point.
Note that \mg already uses a simple grid-based surrogate for the virtual corrections and corrects its error by evaluating the exact virtual amplitude for a fraction of the phase-space points, see Sec.~2.4.3 of Ref.~\cite{Alwall:2014hca}.
In a fixed-order calculation, the same correction can be applied to our neural-network-based surrogates and removes the surrogate bias.
Since the required fraction of exact calls decreases with the surrogate error, the high precision of our surrogates will allow for an even lower fraction of exact virtual amplitude calls.
For unweighted event generation, which is beyond the scope of this fixed-order study, the surrogate bias can instead be removed with a second unweighting step based on the exact amplitudes~\cite{Danziger:2021eeg,Janssen:2023ahv,Herrmann:2025nnz}.
Once the virtual amplitudes are replaced by surrogates, the exact real-emission amplitude in the subtraction region, involving up to four $\Sigma_{ij}$ evaluations, becomes the new bottleneck.
As expected, the acceleration becomes more significant towards higher multiplicities.
Realizing this acceleration in practice requires training \madnis and the surrogates once.

This training is a one-time cost that is not included in Eq.\eqref{eq:speedup} and is amortized over all subsequent runs.
Its dominant part is the generation of the exact training amplitudes, which corresponds to about $5\%$ to $10\%$ of the exact evaluations in one validation run of Fig.~\ref{fig:madnis_observables-4j}.
The network training takes about one GPU-hour.
For a single run, this cost reduces the acceleration, and its impact decreases with the number of runs.
Since our implementation is a proof of concept, we have not optimized the training workflow, and a dedicated implementation will reduce this cost further.

%\clearpage
%%%%%%%%%%%%%%%%%%%%%%%%%%%%%%%%%%%%%%%%%%%%%%%%%%%
\section{Outlook}
\label{sec:outlook}

We have presented a coherent ML framework for subtraction-based NLO calculations, combining amplitude surrogates with neural importance sampling. Virtual corrections are particularly well suited for surrogates, and the corresponding uncertainty-aware precision surrogates are already available. We found that learning the virtual-to-Born ratio performed best without including the integrated subtraction contribution, but incorporating it is straightforward.
In the real-emission subtraction region, the surrogate errors do not cancel and lead to a divergent error in the integral. 
Therefore, we limited our real-emission surrogates to phase-space regions without subtraction. Even then these surrogates are more challenging as the final state contains one additional particle, the range of amplitude values is larger, a ratio-to-Born learning is not obvious, and the FKS-regularized amplitude is not Lorentz invariant.
Their per-point precision is at the percent level and leads to deviations of a few per mille in kinematic distributions.

To complement the surrogates, we have extended \madnis to multi-channel neural importance sampling combined with FKS sectors. These sectors are sampled as additional discrete degrees of freedom. The real-emission surrogates are, correspondingly, FKS-conditioned.
Following a conservative approach, we only used the real-emission surrogates outside the subtraction region and shrank this region by re-optimizing the subtraction thresholds. 
The figure of merit of our study is the acceleration at given Monte Carlo precision.
Here we have found speed gains of a factor of 110 for NLO 3-jet predictions and a factor of 680 for NLO 4-jet predictions, with a surrogate-induced bias below the Monte Carlo uncertainty of our samples.

Our comprehensive surrogate approach makes the entire workflow compatible with GPU parallelization. The one important conceptual question which we did not tackle yet in this study is how to align the subtraction scheme with the strengths and weaknesses of ultrafast amplitude surrogates. 

A natural next step is the extension to hadronic collisions and general LHC processes.
On the sampling side, \madnis already handles hadronic initial states at LO~\cite{Heimel:2023ngj}.
At NLO, initial-state radiation introduces initial-state subtraction terms and degenerate contributions in the FKS scheme.
They depend on the Born-like kinematics and the longitudinal momentum fraction, which can be added as an extra input to the surrogates and an extra dimension of the \madnis sampling.
General LHC processes involve more partonic subprocesses and FKS sectors, which require additional surrogates or a conditioning on the subprocess, and higher multiplicities, where the gain from the virtual surrogates is largest.

%%%%%%%%%%%%%%%%%%%%%%%%%%%%%%%%%%%%%%
\subsubsection*{Code availability}

The code used in this work is publicly available on GitHub as part of the ML for MadGraph organization in the repository
\url{https://github.com/madgraph-ml/madnis-nlo}. The implementation is based on
\texttt{PyTorch} and includes the components required to reproduce the workflows presented in this study.

%%%%%%%%%%%%%%%%%%%%%%%%%%%%%%%%%%%%%%%%%%%%%%%%%%%
\subsection*{Acknowledgements}

We are grateful to Fabio Maltoni and the entire MadGraph team for their continuous support. This work was supported by the Deutsche Forschungsgemeinschaft (DFG, German Research Foundation) under grant 396021762 -- TRR~257 \textsl{Particle Physics Phenomenology after the Higgs Discovery}. This work is supported by the PDR-Weave grant FNRS-DFG numéro T019324F (40020485), and by FRS-FNRS (Belgian National Scientific Research Fund) IISN projects 4.4503.16 (MaxLHC). This research is also supported through the KISS consortium (05D2022) funded by the German Federal Ministry of Research, Technology, and Space BMFTR in the ErUM-Data action plan, the authors acknowledge support by the state of Baden-Würt\-tem\-berg through bwHPC and the German Research Foundation (DFG) through grant no INST 39/963-1 FUGG (bwForCluster NEMO). NE is funded by the Infosys-Cambridge AI Centre. MZ acknowledges financial support by the MUR (Italy), with funds of the European Union (NextGenerationEU), through the PRIN2022 grant 2022EZ3S3F.

\clearpage
%%%%%%%%%%%%%%%%%%%%%%%%%%%%%%%%%%%%%%%%%%%%%%%%%%%
\appendix
\section{Hyperparameters}

%---------------------------------------------------------------------
\begin{table}[h!]
    \setlength{\tabcolsep}{10pt}
    \centering
    \begin{small} \begin{tabular}{l|c}
        \toprule
         Hyperparameter & value\\
         \midrule
         Precision & double\\
         Epochs & 1000 \\
         Batch size & 1024\\
         Optimizer & Adam\\
         Max.\,learning rate & $10^{-3}$\\
         Scheduler & one-cycle\\
         Number of layers & 3 \\
         Hidden features & 128 \\
         Activation function & GELU\\
         \bottomrule
    \end{tabular} \end{small}
    \caption{Hyperparameters for MLP-I architecture over the Born-like phase space.}
    \label{tab:hyperparameter_MLP}
\end{table}
%---------------------------------------------------------------------
%---------------------------------------------------------------------
\begin{table}[h!]
    \setlength{\tabcolsep}{10pt}
    \centering
    \begin{small} \begin{tabular}{l|c}
        \toprule
         Hyperparameter & Value (3j/4j)\\
         \midrule
         Precision & double\\
         Epochs & 2000 \\
         Batch size & 4096\\
         Optimizer & Adam\\
         Max.\,learning rate & $3\times10^{-4}$\\
         Scheduler & cosine annealing\\
         Number of layers & 3 \\
         Hidden features per network & 128/512 \\
         Activation function & GELU\\
         Lin-log threshold & 1e-9\\
         \bottomrule
    \end{tabular} \end{small}
    \caption{Hyperparameters for real-emission surrogates.}
    \label{tab:hyperparameter_real_MLP}
\end{table}
%---------------------------------------------------------------------
%----------------------------------------------------
\begin{table}[h!]
    \setlength{\tabcolsep}{10pt}
    \centering
    \begin{small}
    \begin{tabular}{l@{\hspace{6pt}}|@{\hspace{6pt}}S[table-format=7.0]S[table-format=5.2(4)]}
        \toprule
        \multirow{2}{*}{Hyperparameter} & \multicolumn{2}{c}{Value} \\
        & {$ \Pep \Pem \to \Pu \Pubar \Pg$} & {$ \Pep \Pem \to \Pu \Pubar \Pg \Pg$} \\
        \cmidrule(lr){1-3}
        \vegas bins & 64 & \multicolumn{1}{c}{64} \\
        \vegas batch size & 16384 & \multicolumn{1}{c}{10000} \\
        \vegas training iterations & 15 & \multicolumn{1}{c}{50} \\
        Drawn samples & {$2\,000\,000$} & \multicolumn{1}{c}{$10\,000\,000$} \\
        \bottomrule
    \end{tabular}
    \end{small}
    \caption{Hyperparameters for pure \vegas integration runs.}
    \label{tab:vegas_hyperparams}
\end{table}
%----------------------------------------------------
%----------------------------------------------------
\begin{table}[h!]
    \setlength{\tabcolsep}{10pt}
    \centering
    \begin{small}
    \begin{tabular}{l@{\hspace{6pt}}|@{\hspace{6pt}}S[table-format=7.0]S[table-format=8.0]}
        \toprule
        \multirow{2}{*}{Hyperparameter} & \multicolumn{2}{c}{Value} \\
        & {$ \Pep \Pem \to \Pu \Pubar \Pg$} & {$ \Pep \Pem \to \Pu \Pubar \Pg \Pg$} \\
        \cmidrule(lr){1-3}
        \vegas bins & 64 & 64 \\
        \vegas batch size & 10000 & 10000 \\
        \vegas pretraining iterations & 3 & 10 \\
        \madnis batch size & \multicolumn{1}{l}{\makebox[9.35em][l]{$4\times 256+512$}} & \multicolumn{1}{l}{\makebox[9.6em][c]{$6\times 256+512$}} \\
        Loss & \text{stratified variance} & \text{clipped stratified variance} \\
        \madnis iterations & 10000 & 15000 \\
        Drawn samples & {$2\,000\,000$} & {$50\,000\,000$} \\
        \bottomrule
    \end{tabular}
    \end{small}
    \caption{Hyperparameters for \madnis integration runs.}
    \label{tab:madnis_hyperparams}
\end{table}
%----------------------------------------------------

%%%%%%%%%%%%%%%%%%%%%%%%%%%%%%%%%%%%%%%%%%%%%%%%%%%
\bibliography{tilman,refs}
\end{document}

%% file: incl_settings.tex
\usepackage[utf8]{inputenc} % input encodings (allow UTF-8 input)
\usepackage[T1]{fontenc} 	% selecting font encodings (use 8-bit T1 fonts)
\usepackage[english]{babel} % multilingual support (English language/hyphenation)

\usepackage[bitstream-charter]{mathdesign}
\usepackage{pifont}
\usepackage{geometry} 		% flexible and complete interface to document dimensions
\usepackage{mathtools} 		% math package (fixes various deficiencies of amsmath)
\usepackage{float} 			% floating objects such as figures and tables
\usepackage{graphicx} 		% enhanced support for graphics
\usepackage{tabularx} 		% tabulars with adjustable-width columns
\usepackage{booktabs} 		% professional-quality tables
\usepackage{color, xcolor} 	% foreground and background colour management
\definecolor{Rcolor}{HTML}{E99595}
\definecolor{Gcolor}{HTML}{C5E0B4}
\definecolor{Bcolor}{HTML}{9DC3E6}
\definecolor{Ycolor}{HTML}{FFE699}
\definecolor{Vcolor}{HTML}{CFB4E0}
\usepackage{pdfpages} 		% inclusion of external multi-page PDF documents
\usepackage{extarrows} 		% extra arrows beyond those provided in amsmath
\usepackage{multirow} 		% create tabular cells spanning multiple rows
\usepackage{multicol} 		% intermix single and multiple columns
\usepackage{enumitem} 		% control layout of itemize, enumerate, description
\usepackage{xspace} 		% define commands that appear not to eat spaces
\usepackage{stackrel} 		% enhancement to the \stackrel command
\usepackage{tikz} 			% create PostScript and PDF graphics
\tikzset{
  flow/.style={-Latex, thick},
}
\usetikzlibrary{arrows, shapes.geometric, arrows.meta, shapes, decorations.pathreplacing, fit, patterns, patterns.meta, positioning, calc, fadings}
\usepackage{braket} 		% Dirac bra-ket notation
\usepackage{bm} 			% access bold symbols in maths mode
\usepackage{tensor} 		% typeset tensors (tensor-style super- and subscripts)
\usepackage{slashed} 		% slash through characters (Feynman slash notation)
\usepackage{siunitx} 		% SI units package (typesetting values with units)
\usepackage{lastpage} 		% reference last page
\usepackage{cite} 			% improved citation handling
\usepackage[normalem]{ulem} % package for underlining
\usepackage{fontawesome} 	% access to web-related icons
\usepackage{tocloft} 		% control over the typography of the TOC, LOF, and LOT
\usepackage{titlesec} 		% interface to sectioning commands (various title styles)
\usepackage{doi} 			% create correct hyperlinks for DOI numbers
\usepackage{hyperref} 		% hypertext links (handel cross-referencing commands)
\usepackage[most]{tcolorbox} 					% coloured and framed text boxes
\usepackage[nameinlink, capitalize]{cleveref} 	% intelligent cross-referencing
\usepackage[nottoc, notlot, notlof]{tocbibind} 	% add references/index/contents to TOC
\usepackage[ruled, vlined]{algorithm2e} 		% floating algorithm environment
\usepackage{makecell}
\usepackage[makeroom]{cancel}
\usepackage{feynmf}
\usepackage{accents}

\usepackage[colorinlistoftodos]{todonotes}
\makeatletter
\def\BState{\State\hskip-\ALG@thistlm}
\makeatother

\makeatletter
\@ifundefined{pdfoutput}{}{\DeclareGraphicsRule{*}{mps}{*}{}}
\makeatother

\makeatletter
\DeclareRobustCommand*{\bfseries}{%
   \not@math@alphabet\bfseries\mathbf
   \fontseries\bfdefault\selectfont
   \boldmath
}
\makeatother

\hypersetup{
	pdftitle={MadNIS at NLO},
	pdfauthor={De Crescenzo et al.},
	colorlinks=true, 			% false: boxed links, true: colored links
	linkcolor={red!50!black}, 	% color of internal links (sections, pages, etc.)
	citecolor={blue!50!black}, 	% color of citation links (links to bibliography)
	urlcolor={blue!80!black} 	% color of URL links (external links)
} 
\DeclareSymbolFont{usualmathcal}{OMS}{cmsy}{m}{n}
\DeclareSymbolFontAlphabet{\mathcal}{usualmathcal}

\SetArgSty{textnormal}
\SetKwComment{Comment}{{\small\#}~}{}
\SetCommentSty{mycommfont}

\setitemize{itemsep=0pt, parsep=0pt} 				% adjust itemize environment
\setenumerate{itemsep=0pt, parsep=0pt} 				% adjust enumerate environment
\setitemize{itemsep=2pt,topsep=2pt,parsep=0pt,partopsep=0pt,leftmargin=*}
\setenumerate{itemsep=0pt,topsep=2pt,parsep=0pt,partopsep=0pt,labelindent=3pt,leftmargin=*}
\newlist{todolist}{itemize}{2}
\setlist[todolist]{label=$\square$}
\usepackage{pifont}
\usepackage{diagbox} % in your preamble

%% file: incl_shortcuts.tex
\definecolor{red_cb}{HTML}{e41a1c}
\definecolor{blue_cb}{HTML}{377eb8}
\definecolor{green_cb}{HTML}{4daf4a}
\definecolor{purple_cb}{HTML}{984ea3}
\definecolor{orange_cb}{HTML}{ff7f00}

\definecolor{EmeraldGreen}{HTML}{1ea78d}
\definecolor{EnglishRed}{HTML}{b02427}
\hypersetup{colorlinks=true,urlcolor=EmeraldGreen,citecolor=EmeraldGreen,linkcolor=EnglishRed}

\newcommand{\eg}{\text{e.g.}\;}
\newcommand{\ie}{\text{i.e.}\;}

\newcommand{\mwith}{\text{with}}
\newcommand{\mand}{\text{and}}
\newcommand{\mfor}{\text{for}}

\newcommand{\qqquad}{\qquad\quad}
\newcommand{\qqqquad}{\qquad\qquad}

\def\dd{\text{d}}

\newcommand\one{\leavevmode\hbox{\small1\normalsize\kern-.33em1}}
\newcommand{\loss}{\mathcal{L}} 	% loss value

\newcommand{\mata}{\mathcal{A}}
\newcommand{\pmother}{p_{j}}
\newcommand{\psister}{\tilde{p}_{j}}
\newcommand{\pdaughter}{\tilde{p}_{i}}

\newcommand{\vegas}{\texttt{VEGAS}\xspace}

\newcommand{\madspace}{\texttt{MadSpace}\xspace}
\newcommand{\mg}{\texttt{MG5aMC}\xspace}
\newcommand{\pythia}{\texttt{Pythia}\xspace}

\newcommand{\sherpa}{\texttt{Sherpa}\xspace}
\newcommand{\herwig}{\texttt{Herwig}\xspace}
\newcommand{\madnis}{\texttt{MadNIS}\xspace}

\newcommand{\arXiv}[2][]{%
	\ifthenelse{\equal{#1}{}}%
	{\href{http://arxiv.org/abs/#2}{arXiv:#2}}%
	{\href{http://arxiv.org/abs/#2}{arXiv:#2~[#1]}}}

\newcommand{\tev}{\text{TeV}}

\def\slashchar#1{\setbox0=\hbox{$#1$}           % set a box for #1
   \dimen0=\wd0                                 % and get its size
   \setbox1=\hbox{/} \dimen1=\wd1               % get size of /
   \ifdim\dimen0>\dimen1                        % #1 is bigger
      \rlap{\hbox to \dimen0{\hfil/\hfil}}      % so center / in box
      #1                                        % and print #1
   \else                                        % / is bigger
      \rlap{\hbox to \dimen1{\hfil$#1$\hfil}}   % so center #1
      /                                         % and print /
   \fi}

\newcommand{\tikznode}[2]{%
\ifmmode%
\tikz[remember picture,baseline=(#1.base),inner sep=0pt] \node (#1) {$#2$};%
\else
\tikz[remember picture,baseline=(#1.base),inner sep=0pt] \node (#1) {#2};%
\fi}

\newcommand{\ms}[1]{\ensuremath{#1}\xspace}
\newcommand{\mr}[1]{\ensuremath{\mathrm{#1}}\xspace}

\newcommand{\Pq}{\ms{q}}
\newcommand{\Pqbar}{\ms{\bar q}}

\newcommand{\Pg}{\mr{g}}

\newcommand{\Pep}{\mr{e}^{+}}
\newcommand{\Pem}{\mr{e}^{-}}

\newcommand{\Pd}{\mr{d}}
\newcommand{\Pu}{\mr{u}}
\newcommand{\Ps}{\mr{s}}
\newcommand{\Pc}{\mr{c}}

\newcommand{\Pubar}{\mr{\bar u}}
\newcommand{\Pdbar}{\mr{\bar d}}
\newcommand{\Psbar}{\mr{\bar s}}
\newcommand{\Pcbar}{\mr{\bar c}}

%% file: figs/tikz/fks.tex
\definecolor{NNviolet}{RGB}{170,60,255} % bright violet
\begin{tikzpicture}[font=\small]

\node (Sample_FKS_sec) [rectangle, rounded corners, fill=Gcolor,
  minimum width=5.75cm, minimum height=1.2cm, align=center, anchor=west]
{Sample $ij \sim p_{\theta_k}(ij)$};

\node (NNFrame) [
  draw=NNviolet,
  line width=2pt,
  draw opacity=.75,
  rounded corners,
  inner sep=0.1pt,
  fit=(Sample_FKS_sec)
] {};

\node (Sample_kinematics) [rectangle, rounded corners, fill=Gcolor,
  minimum width=5.75cm, minimum height=1.2cm, below=1.75cm of Sample_FKS_sec.west,
  align=center, anchor=west]
{Sample $(x_\text{B}, x_\text{rad}) \sim q_{\theta_k}(x_\text{B}, x_\text{rad} | ij)$};

\node (NNFrame2) [
  draw=NNviolet,
  line width=2pt,
  draw opacity=.75,
  rounded corners,
  inner sep=0.1pt,
  fit=(Sample_kinematics)
] {};

\draw[flow, shorten >=0.4mm, shorten <=0.4mm] (Sample_FKS_sec) -- (Sample_kinematics);

\node (MapBorn) [rectangle, rounded corners, fill=Bcolor, minimum width=5.75cm, minimum height=1.2cm, align=center, below=1.75cm of Sample_kinematics.west, anchor=west] {Map $x_\text{B}\rightarrow \Phi_n$};

\draw[flow, shorten <=0.4mm] (Sample_kinematics) -- (MapBorn);

\node (MapRad) [rectangle, rounded corners, fill=Bcolor, minimum width=5.75cm, minimum height=1.2cm, align=center, right=0.75cm of Sample_kinematics.east] {Map $x_\text{rad}\rightarrow \{\xi_i,y_{ij},\varphi_i\}$};

\draw[flow, shorten <=0.4mm] (Sample_kinematics) -- (MapRad);

\node (ConstructNp1) [rectangle, rounded corners, fill=Ycolor, minimum width=5.75cm, minimum height=1.2cm, align=center, below=1.75cm of MapRad.east, anchor=east] {Construct $(\Phi_{n+1}^\text{hard}, \Phi_{n+1}^\text{soft}, \Phi_{n+1}^\text{coll}, \Phi_{n+1}^\text{soft-coll})$ \\ from $\{ij, \Phi_n, \xi_i,y_{ij},\varphi_i\}$};

\draw[flow] (MapRad) -- (ConstructNp1);
\draw[flow] (MapBorn.east) -- (ConstructNp1.west |- MapBorn.east);

\coordinate (MidBC) at ($(MapBorn)!0.5!(ConstructNp1)$);
\node (Cuts) [rectangle, rounded corners, fill=Ycolor,
  minimum width=12.25cm, minimum height=1.2cm, align=center,
  below=1.15cm of MidBC]
{Apply $n$-body cuts to $(\Phi_n, \Phi_{n+1}^\text{hard}, \Phi_{n+1}^\text{soft}, \Phi_{n+1}^\text{coll}, \Phi_{n+1}^\text{soft-coll})$};

\draw[flow] (MapBorn.south) -- (MapBorn.south |- Cuts.north);
\draw[flow] (ConstructNp1.south) -- (ConstructNp1.south |- Cuts.north);

\node (Evaluation nbody) [rectangle, rounded corners, fill=Rcolor,
  minimum width=5.75cm, minimum height=1.2cm, align=center, below=1.75cm of Cuts.west, anchor=west]
{Evaluate $\mata^\text{B}(\Phi_{n})$, $\mata^\text{V}(\Phi_{n})$, $\mata^\text{I}(\Phi_{n})$};
\node (Evaluation np1body) [rectangle, rounded corners, fill=Rcolor,
  minimum width=5.75cm, minimum height=1.2cm, align=center,
      below=1.75cm of Cuts.east, anchor=east]
{Evaluate $\mata_{ij}^{\text{R}-\text{S}}(\Phi_{n+1})$};

% Some manual braces
\coordinate (bL1) at ($(Cuts.south west)+(5.65cm,9pt)$);
\coordinate (bL2) at ($(Cuts.south west)+(6.15cm,9pt)$);

\coordinate (bR1) at ($(Cuts.south west)+(6.2cm,9pt)$);
\coordinate (bR2) at ($(Cuts.south west)+(9.75cm,9pt)$);

\draw[line width=0.9pt, decorate, decoration={brace,mirror,amplitude=4pt}]
  (bL1) -- (bL2) coordinate[midway] (braceB);

\draw[line width=0.9pt, decorate, decoration={brace,mirror,amplitude=4pt}]
  (bR1) -- (bR2) coordinate[midway] (braceR);

\coordinate (startB) at ($(braceB)+(0,-6pt)$);
\coordinate (startR) at ($(braceR)+(0,-6pt)$);

\coordinate (xB) at (MapBorn.south |- Cuts.north);
\coordinate (xR) at (ConstructNp1.south |- Cuts.north);

% choose a routing y-level above the evaluation nodes (tune the +6mm if you want)
\coordinate (routeY) at ($(Evaluation nbody.north)+(0,4mm)$);

% down -> sideways -> down (ending at same x as the Cuts-arrows)
\draw[flow, shorten >=0.4mm] (startB) -- (startB |- routeY) -| (xB |- routeY) -- (xB |- Evaluation nbody.north);

\draw[flow, shorten >=0.4mm] (startR) -- (startR |- routeY) -| (xR |- routeY) -- (xR |- Evaluation np1body.north);

% fit everything under one big rectangle and write on top channel k
\node (ChannelFrame) [
  draw,
  rounded corners,
  dashed,
  line width=1.2pt,
  fill=none,
  inner sep=5pt,
  fit=(Sample_FKS_sec) (Sample_kinematics) (MapBorn) (MapRad)
      (ConstructNp1) (Cuts) (Evaluation nbody) (Evaluation np1body)
] {};

\node[anchor=south east, inner sep=2pt]
  at ($(ChannelFrame.north east)+(0pt,1pt)$) {channel $k$};

\node (NNFrame3) [
  draw=NNviolet,
  line width=2pt,
  draw opacity=.75,
  rounded corners,
  inner sep=0.1pt,
  fit=(Evaluation nbody)
] {};

\node (NNFrame4) [
  draw=NNviolet,
  line width=2pt,
  draw opacity=.75,
  rounded corners,
  inner sep=0.1pt,
  fit=(Evaluation np1body)
] {};

\end{tikzpicture}